\documentclass[12pt]{iopart}
\usepackage{psfrag,epsfig,iopams}
\usepackage{graphicx}
\usepackage{color}
\usepackage{hyperref,cite}
\usepackage{units}
\usepackage{stackrel,appendix}

\newcommand*{\TRANS}{^{\mkern-1.5mu\mathsf{T}}}
\newcommand*{\VEC}[1]{\boldsymbol{#1}}
\newcommand*{\TENSOR}[1]{\mathsf{#1}}
\newcommand*{\OP}[1]{{\cal{#1}}}
\newcommand*{\kB}{k_{\mathrm{B}}}

\renewcommand{\d}{\mathrm{d}}	

\newcommand*{\fnc}{\VEC{f}_{\mathrm{nc}}}
\newcommand*{\g}{\bar\gamma}
\newcommand*{\tf}{{t_{\mathrm{f}}}}
\newcommand*{\tb}{{\tau_{\mathrm{b}}}}
\newcommand*{\F}{{\cal F}}

\newcommand*{\Traj}[1]{\overline{#1}}

\newcommand{\changed}[1]{{\color{black}#1}}

\begin{document}
\title{Functionals in stochastic thermodynamics:
how to interpret stochastic integrals}
\author{Stefano Bo\footnote{Present address: Max Planck Institute for the Physics of Complex Systems, N{\"o}thnitzer Str. 38, DE-01187 Dresden, Germany.}, Soon Hoe Lim and Ralf Eichhorn}
\address{Nordita, Royal Institute of Technology and Stockholm University, Roslagstullsbacken 23, SE-106 91 Stockholm, Sweden}
\eads{stefabo@pks.mpg.de}

\begin{abstract}
In stochastic thermodynamics standard concepts from macroscopic thermodynamics, such
as heat, work, and entropy production, are generalized to small fluctuating systems
by defining them on a trajectory-wise level.
In Langevin systems with continuous state-space
such definitions involve stochastic integrals along system trajectories,
whose specific values 
 depend on the discretization rule used
to evaluate them (i.e.\ the ``interpretation'' of the noise terms in the integral).
Via a systematic mathematical investigation of this apparent dilemma,
we corroborate the widely used standard interpretation of heat- and work-like
functionals as Stratonovich integrals. We furthermore recapitulate the 
anomalies that are
known to occur for entropy production in the presence of temperature gradients. 
\end{abstract}

\date{\today}
\maketitle

\section{Introduction}
\label{sec:intro}
In stochastic thermodynamics standard concepts of macroscopic thermodynamics are
generalized to cover the non-equilibrium properties of small systems
\cite{jarzynski2011equalities,seifert2012stochastic}.
Instead of being obtained as ensemble averages,
thermodynamic quantities such as heat, work, and entropy changes are defined
on the level of individual trajectories \cite{van2015ensemble},
allowing for a refined formulation of the thermodynamic laws.
For instance, in an overdamped, $d$-dimensional Langevin system
\cite{vanKampen1992stochastic,mazo2002brownian,philipse2018brownian} with
coordinate $\VEC{x}_t$ at time $t$,
\begin{equation}
\label{eq:overdampedLE}
\gamma\dot{\VEC{x}_t} = -\nabla_{\VEC{x}_t} U(\VEC{x}_t,t) + \fnc(\VEC{x}_t,t) + \sqrt{2\kB T \gamma} \, \VEC{\xi}_t
\, ,
\end{equation}
the first law for a trajectory $\Traj{\VEC{x}}=\{\VEC{x}_t\}_{t=0}^\tf$ evolving from
$\VEC{x}_0$ at time $t=0$ to $\VEC{x}_{\mathrm{f}}$ at time $t=\tf$ reads
\begin{equation}
\label{eq:1stLaw}
\Delta U[\Traj{\VEC{x}}] = Q[\Traj{\VEC{x}}] + W[\Traj{\VEC{x}}]
\, ,
\end{equation}
with the definitions
\numparts
\begin{eqnarray}
\Delta U[\Traj{\VEC{x}}]
&=& U(\VEC{x}_{\mathrm{f}},\tf) - U(\VEC{x}_0,0)
\label{eq:DeltaU}
\, , \\[1ex]
Q[\Traj{\VEC{x}}]
&=& \int_0^\tf \left( -\gamma\dot{\VEC{x}}_t+\sqrt{2\kB T\gamma}\,\VEC{\xi}_t \right) \circ \d\VEC{x}_t
\nonumber \\
&=& \int_0^\tf \Big( \nabla_{\VEC{x}_t} U(\VEC{x}_t,t) - \fnc(\VEC{x}_t,t) \Big)\circ\d\VEC{x}_t
\label{eq:Q}
\, , \\[1ex]
W[\Traj{\VEC{x}}]
&=& \int_0^\tf \left(
	\frac{\partial U(\VEC{x}_t,t)}{\partial t}\,\mathrm{d}t + \fnc(\VEC{x}_t,t) \circ \d\VEC{x}_t
  \right)
\label{eq:W}
\, .
\end{eqnarray}
\endnumparts
In \eref{eq:overdampedLE}, the deterministic force exerted on the system has a
conservative component $-\nabla_{\VEC{x}} U(\VEC{x},t)$ due to an external potential $U(\VEC{x},t)$,
and a non-conservative component $\fnc(\VEC{x},t)$.
The dot in $\dot{\VEC{x}}_t$ denotes the total time derivative,
$\gamma$ is the friction coefficient, $\kB$ Boltzmann's constant,
and $T$ the temperature of the thermal bath.
The thermal fluctuations $\VEC{\xi}_t$ are modeled by Gaussian white noise sources,
which are unbiased, mutually independent and delta-correlated in time.
The $\circ$ symbol in \eref{eq:Q} and \eref{eq:W} denotes the Stratonovich product,
which implies mid-point discretization. The second line in \eref{eq:Q} is simply
obtained by exploiting the equation of motion \eref{eq:overdampedLE}.
A generic example for a physical system described by the Langevin equation \eref{eq:overdampedLE}
is a colloidal Brownian particle suspended in an aqueous solution at room temperature.

The definitions in \eref{eq:DeltaU}, \eref{eq:Q} and \eref{eq:W}
are physically motivated.
The change in \emph{internal energy} $\Delta U[\Traj{\VEC{x}}]$ is given by the
difference of the system's potential energy at the beginning and the end of the
trajectory $\Traj{\VEC{x}}$, and thus is dependent only on initial and final state.
The \emph{heat} $Q[\Traj{\VEC{x}}]$ absorbed by the system results from the energy
exchange due to the frictional force $-\gamma\dot{\VEC{x}}_t$ and fluctuating force
$\sqrt{2\kB T\gamma}\,\VEC{\xi}_t$
exerted by the heat bath on the system \cite{sekimoto2010stochastic},
integrated over all displacements $\mathrm{d}\VEC{x}_t$
along the trajectory $\Traj{\VEC{x}}$.
The \emph{work} $W[\Traj{\VEC{x}}]$ performed on the system by an ``external agent''
\cite{sekimoto2010stochastic} results from externally controlled changes
of the potential energy $\frac{\partial U(\VEC{x}_t,t)}{\partial t}\,\d t$
and from non-conservative forces $\fnc(\VEC{x}_t,t) \circ \mathrm{d}\VEC{x}_t$,
acting during the motion along $\Traj{\VEC{x}}$.
Therefore, both heat and work are functionals along the trajectory $\Traj{\VEC{x}}$.
As argued in \cite{sekimoto2010stochastic}, the consistency of these physical
interpretations with
the energy balance \eref{eq:1stLaw} is based on interpreting the products in
\eref{eq:Q} and \eref{eq:W}
in the \emph{Stratonovich} sense (marked by $\circ$), because then the usual
rules of calculus can be applied when computing 
differentials, such as
$\d U(\VEC{x}_t,t) = \nabla_{\VEC{x}_t} U(\VEC{x}_t,t)\circ\d\VEC{x}_t
+\frac{\partial U(\VEC{x}_t,t)}{\partial t}\,\mathrm{d}t$
for the infinitesimal change in internal energy.

With similar reasoning
functionals along the trajectory $\Traj{\VEC{x}}$,
which are not directly obtained from energy balance arguments,
are defined employing standard rules of calculus.
The most prominent example is the \emph{entropy production in the thermal bath}
\cite{jarzynski2011equalities,seifert2012stochastic}
\numparts
\begin{equation}
\label{eq:DeltaSdef}
\Delta S^{\mathrm{env}}[\Traj{\VEC{x}}]
= -\int_0^\tf \frac{\d Q_t}{T}
\end{equation}
Using the definition
$\d Q_t=\left( -\gamma\dot{\VEC{x}}_t+\sqrt{2\kB T\gamma}\,\VEC{\xi}_t \right) \circ \d\VEC{x}_t$
for the infinitesimal heat exchange over a displacement $\d\VEC{x}_t$ in accordance with
\eref{eq:Q}, and replacing $-\gamma\dot{\VEC{x}}_t+\sqrt{2\kB T\gamma}\,\VEC{\xi}_t$ with
the balancing external forces from the Langevin equation \eref{eq:overdampedLE},
we find
\begin{equation}
\label{eq:DeltaS}
\Delta S^{\mathrm{env}}[\Traj{\VEC{x}}]
= \int_0^\tf \frac{1}{T}\Big( -\nabla U(\VEC{x}_t,t) + \fnc(\VEC{x}_t,t) \Big)\circ\d\VEC{x}_t
\, .
\end{equation}
\endnumparts
Following this common wisdom we thus end up with a Stratonovich interpretation of the
stochastic integral in \eref{eq:DeltaS}.
Another strong indication for the Stratonovich interpretation in \eref{eq:DeltaS}
being adequate comes from the fact that then $\Delta S^{\mathrm{env}}[\Traj{\VEC{x}}]$ can be
interpreted as a measure for the (ir-)reversibility of the path $\Traj{\VEC{x}}$.
It has been shown \cite{kurchan1998fluctuation,seifert2012stochastic,chetrite2008fluctuation} that the log-ratio of probabilities for
the trajectory $\Traj{\VEC{x}}$ being generated by \eref{eq:overdampedLE}
in forward time versus the same
trajectory being traced out backwards under a time-reversed dynamics
is exactly given by \eref{eq:DeltaS} as a Stratonovich integral.

From a more mathematical viewpoint the above arguments may not be sufficiently
convincing to
resolve the fundamental ambiguity of how to ``correctly'' define the integrals
in \eref{eq:Q}, \eref{eq:W} along the stochastic trajectory $\Traj{\VEC{x}}$:
It\^o interpretation, Stratonovich or even something else?
This problem becomes even more apparent if the
friction $\gamma$ or the temperature $T$
are functions of the position $\VEC{x}$,
because then the associated multiplicative noise term in the
Langevin equation may have to be interpreted in either It\^o or
anti-It\^o (also called H\"anggi-Klimontovich \cite{hanggi1982stochastic,klimontovich1994nonlinear})
convention, depending on which of the physical quantities, $\gamma$ or $T$,
actually varies in space
\cite{sancho1982adiabatic,jayannavar1995macroscopic,stolovitzky1998non,yang2013brownian,hottovy2012noise,hottovy2015smoluchowski}
(see also our derivation in Section \ref{sec:limits} below).
The identification of the force contributions to heat
and work is then far less obvious already on the level of the equation of motion
\cite{sekimoto2010stochastic}.
Put differently, it is of fundamental interest 
to verify, via a systematic approach, whether the definitions
\eref{eq:DeltaU}, \eref{eq:Q}, \eref{eq:W} and \eref{eq:DeltaS}
are indeed the correct ones or whether they need modifications,
in particular for more general cases than
the simple Langevin equation \eref{eq:overdampedLE} with additive noise.

We provide such a mathematical analysis with the present contribution,
\changed{treating dynamics and thermodynamics under a common approach
along the lines we have employed, together with collaborators,
in earlier works
\cite{celani2012anomalous,bo2013white,bo2014entropy,aurell2016diffusion,marino2016entropy,bo2017multiple,Lim2018}.
In addition to deriving new results concerning functionals,
we recover and systematize
a number of related findings known from the literature, as we
point out throughout the paper in connection with the relevant results.
From this perspective our present work can also be seen as being of review-like
nature, though certainly not exhaustive
(e.g.\ not covering the most general cases).
}

The ambiguity in interpreting products and integrals
involving the stochastic process $\VEC{x}_t$ generated by
the Langevin equation \eref{eq:overdampedLE} is related to two
asymptotic limits that have implicitly been performed before writing
down \eref{eq:overdampedLE}:
(i) the limit of vanishing correlation time of the thermal noise
(white-noise limit), and
(ii) the limit of vanishing equilibration time for the velocity
degrees of freedom (small-mass limit \cite{purcell1977life}).
We therefore start our analysis from the so-called generalized
Langevin equation (GLE) \cite{hanggi1997generalized}, i.e.\ a generalized form of the
Langevin equation which possesses finite memory times and  noise correlation times,
and includes velocity degrees of freedom with
finite relaxation time (see Sec.~\ref{sec:GLE}).
We then perform the aforementioned asymptotic limits in a
systematic way not only
for the Langevin equation of motion itself, but also for general
functionals along trajectories generated by the GLE.

Much has been written about the noise interpretation related to the 
white-noise limit and the small-mass limit in stochastic differential equations (SDEs), 
with the two limits usually studied separately
\cite{sancho1982adiabatic,hanggi1994colored,jayannavar1995macroscopic,yang2013brownian,kupferman2004ito,hottovy2012noise,hottovy2015smoluchowski,pavliotis2008multiscale,stolovitzky1998non,moon2014interpretation,wong1965convergence,bo2013white},
and, more recently, about the behavior of stochastic thermodynamic quantities in the small-mass limit
\cite{celani2012anomalous,spinney2012entropy,bo2013entropic,kawaguchi2013fluctuation,bo2014entropy,ge2014time,nakayama2015invariance,ford2015stochastic,marino2016entropy,bo2017multiple,PhysRevE.98.052105,ge2018anomalous,miyazaki2018entropy,birrell2018entropy}.
\changed{As mentioned above, we will recover several of these previous results
along our analysis,}
and present them in a single framework.

\section{Model}
\subsection{The generalized Langevin equation (GLE)}
\label{sec:GLE}
The generalized form of the Langevin equation which we take as a starting point
of our analysis reads \cite{Lim2018}
\begin{eqnarray}
m\ddot{\VEC{x}}_t
= \VEC{f}(\VEC{x}_t,t)
- \g\,\TENSOR{g}(\VEC{x}_t) \int_0^t \d s \, {\kappa}(t-s) \TENSOR{h}(\VEC{x}_s)\dot{\VEC{x}}_s
\nonumber \\ \qquad\qquad\qquad \mbox{}
+ \sqrt{2\kB T(\VEC{x}_t,t)\g}\,\TENSOR{\sigma}(\VEC{x}_t)\VEC{\eta}_t
\label{eq:GLE}
\, .
\end{eqnarray}
It describes the dynamics of a Brownian particle (with mass $m$) with position $\VEC{x}_t$
and velocity $\VEC{v}_t=\dot{\VEC{x}}_t$
(in $d=1,2\mbox{ or }3$ dimensions) at time $t$,
which starts out from $\VEC{x}_0$ with velocity $\VEC{v}_0$ at time $t=0$.
The particle moves in a force field
$\VEC{f}(\VEC{x},t)=-\nabla_{\VEC{x}} U(\VEC{x},t)+\fnc(\VEC{x},t)$ and interacts with its environment.
This interaction leads to frictional and fluctuating forces,
which are modeled by the second and third term, respectively, on the right-hand side of
the generalized Langevin equation \eref{eq:GLE}.

The functions $\TENSOR{g}(\VEC{x})$, $\TENSOR{h}(\VEC{x})$, and $\TENSOR{\sigma}(\VEC{x})$
are all matrix-valued, i.e.\ they are functions from $\mathbb{R}^{d}$ to $\mathbb{R}^{d\times d}$.
The viscous friction term (second term on the right-hand side in \eref{eq:GLE}),
involving an integral over the particle's past velocities
with the kernel $\kappa(t-s)$, describes state-dependent dissipation
and comprises the back-action effects of the environment up to the current time $t$.
For the memory function $\kappa(t)$ we here consider the case
\begin{equation}
\label{eq:kappa}
\TENSOR{\kappa}(t) = \frac{1}{\tb}\, e^{-t/\tb}
\, ,
\end{equation}
with $\tb$ being the characteristic time scale of memory
effects induced by the thermal bath.
The typical magnitude of the viscous friction forces is specified by the factor $\g$,
i.e.\ $\TENSOR{g}$, $\TENSOR{h}$ and $\TENSOR{\sigma}$ are chosen to be dimensionless.

The noise term $\sqrt{2\kB T(\VEC{x}_t,t)\g}\,\TENSOR{\sigma}(\VEC{x}_t)\VEC{\eta}_t$ in \eref{eq:GLE}
also accounts for past and present interactions
of the particle with the environment, where $\VEC{\eta}_t$ is a mean-zero, stationary
colored noise process.
Note that even though they depend on position, $\VEC{x}_t$, there is no ambiguity related to the
interpretation of these products because
\changed{$\VEC{x}_t$ is the second integral of the noise process
and therefore is sufficiently regular}
\footnote{\changed{Note that the noise process here happens to be colored so that in effect
$\VEC{x}_t$ is the third integral of a white noise.}}.
We choose the $\VEC{\eta}_t$ to be Gaussian with time-exponential correlations
\begin{equation}
\label{eq:etaietaj}
\langle \eta^i_t\eta^j_s \rangle = \frac{\delta_{ij}}{2\tb} \, e^{-|t-s|/\tb}
\, ,
\end{equation}
where the correlation time scale $\tb$ is identical to the characteristic time scale of the
memory kernel \eref{eq:kappa} of the viscous friction, because
both memory and colored noise effects are governed
by the intrinsic time scale of the bath.
This is equivalent to generating the $\VEC{\eta}_t$ by the
stationary Ornstein-Uhlenbeck process, which is the solution to the SDE:
\begin{equation}
\label{eq:etadot}
\dot{\VEC{\eta}}_t = -\frac{1}{\tb}\VEC{\eta}_t + \frac{1}{\tb}\VEC{\xi}_t
\, ,
\end{equation}
where $\VEC{\xi}_t$ is a mean-zero Gaussian white noise source
with covariance $\langle \xi^i_t\xi^j_s \rangle=\delta_{ij}\delta(t-s)$.
The typical scale of the colored noise strength is set by the space- and time-dependent
parameter $T(\VEC{x},t)$.

Note that we allow the functions $\TENSOR{g}(\VEC{x})$, $\TENSOR{h}(\VEC{x})$,
and $\TENSOR{\sigma}(\VEC{x})$ to be unrelated to each other in general.
Such a general setting does not necessarily fulfill the
fluctuation-dissipation relation (in current notation) \cite{kubo1995statistical},
\begin{equation}
\label{eq:FDT}
2\kB T \g \left\langle \left[ \TENSOR{\sigma}(\VEC{x}_t)\VEC{\eta}_t \right]_i
					\left[ \TENSOR{\sigma}(\VEC{x}_s)\VEC{\eta}_s \right]_j \right\rangle
= \kB T \g \left[ \TENSOR{g}(\VEC{x}_t) \kappa(t-s) \TENSOR{h}(\VEC{x}_s) \right]_{ij}
\, ,
\end{equation}
and thus corresponds to the Brownian particle being in contact with
an inherently \emph{non-equilibrium} environment.
We immediately see, however, that the fluctuation-dissipation relation \eref{eq:FDT} is
valid when $\TENSOR{g}=\TENSOR{h}\TRANS=\TENSOR{\sigma}$.
Then, the latter tensors are typically
the ``square root'' $\TENSOR{\gamma}^{1/2}$ of the (symmetric) hydrodynamic
friction tensor $\TENSOR{\gamma}$ (here measured in units of $\g$).
Furthermore, the parameter $T$ corresponds to the temperature of the environment.
In the case $T=T(\VEC{x},t)$, this environment is
locally ``around position $\VEC{x}$'' and at any given time $t$
an equilibrium heat bath at definite temperature.

We finally point out that in introducing the factors $\g$ and $\sqrt{2\kB T \g}$ in \eref{eq:GLE}
we use a somewhat untypical representation of the GLE, compare for instance with \cite{hanggi1997generalized,Lim2018}.
However, this form \eref{eq:GLE} makes the connection to the overdamped Langevin equation \eref{eq:overdampedLE}
more explicit and thus the generalization of the functionals \eref{eq:Q}, \eref{eq:W} 
to the GLE case more evident, see Sec.~\ref{sec:Gfunctionals}.
We also remark that the form \eref{eq:GLE} of the GLE is more general than the generic GLE
which is typically obtained from microscopic Hamiltonian models for a small system interacting
with a heat bath \cite{hanggi1997generalized,zwanzig1973nonlinear}
(see, for instance, Appendix A in \cite{Lim2018} for a derivation).
In particular, in these systems temperature would be a space- and time-homogenous constant.
While space-dependent friction coefficients are known to be related to nonlinear
system-bath couplings \cite{hanggi1997generalized}, the space- (and time-) dependent temperature
is a generalization we introduce ``by hand'' under the following assumption:
the temperature field $T(\VEC{x},t)$ varies sufficiently slowly such that the particle
at position $\VEC{x}$ and time $t$ is in contact with a locally
well-defined thermal bath, which is homogeneous on mesoscopic scales, and that any
memory effects from that local bath on time scales $\tb$ die out before the
particle changes position into an ``adjacent bath'' at (slightly) different temperature,
or before the bath-temperature changes significantly in time.
The latter assumption is justified because we consider $\tb$ to be
the (by far) fastest time scale in the system (see discussion in Sec.~\ref{sec:formulation}). 

\subsection{The generalized functionals in stochastic thermodynamics}
\label{sec:Gfunctionals}
With the physical interpretation of the various terms in \eref{eq:GLE}
as external force, (generalized) viscous friction and
thermal fluctuations, we
can now define heat and work
\footnote{Note that we here use 
heat and work as names for the functionals
\eref{eq:QGLE} and \eref{eq:WGLE}
even in the general case in which the 
fluctuation-dissipation relation \eref{eq:FDT} is violated.
From a strict physical viewpoint this is not completely precise,
in particular for the heat dissipated in the environment,
which is a physically reasonable concept only if that environment is
an actual ``heat bath'' at thermal equilibrium,
i.e.~if the fluctuation-dissipation theorem is valid.}
along the trajectory
$(\Traj{\VEC{x}},\Traj{\VEC{v}})=\{(\VEC{x}_t,\VEC{v}_t)\}_{t=0}^\tf$
generated by \eref{eq:GLE}
in complete analogy to our discussion in
Sec.~\ref{sec:intro},
\numparts
\begin{eqnarray}
Q[\Traj{\VEC{x}},\Traj{\VEC{v}}]
& = & \int_0^\tf \Bigg(
	- \g\,\TENSOR{g}(\VEC{x}_t) \int_0^t \d s \, {\kappa}(t-s) \TENSOR{h}(\VEC{x}_s){\VEC{x}}_s
\nonumber \\
&& \qquad\qquad\qquad\quad \mbox{}
+ \sqrt{2\kB T(\VEC{x}_t,t)\g}\,\TENSOR{\sigma}(\VEC{x}_t)\VEC{\eta}_t
\Bigg) \cdot \d\VEC{x}_t
\nonumber \\[1ex]
& = & \frac{m}{2} \left( \VEC{v}_{\mathrm{f}}^2 - \VEC{v}_0^2 \right)
	-\int_0^\tf \VEC{f}(\VEC{x}_t,t) \cdot \d\VEC{x}_t
\label{eq:QGLE}
\, , \\[2ex]
W[\Traj{\VEC{x}},\Traj{\VEC{v}}]
& = & \int_0^\tf \left(
	\frac{\partial U(\VEC{x}_t,t)}{\partial t}\,\mathrm{d}t + \fnc(\VEC{x}_t,t) \cdot \d\VEC{x}_t
  \right)
\label{eq:WGLE}
\, .
\end{eqnarray}
\endnumparts
Similarly, the total internal energy for the GLE \eref{eq:GLE} is given as
the sum of kinetic and potential energy,
\begin{equation}
E(\VEC{x},\VEC{v},t) = \frac{m}{2}\VEC{v}^2 + U(\VEC{x},t)
\, .
\end{equation}
We can now verify easily that heat and work together with the change in internal energy
$\Delta E[\Traj{\VEC{x}},\Traj{\VEC{v}}]=E(\VEC{x}_{\mathrm{f}},\VEC{v}_{\mathrm{f}},\tf)-E(\VEC{x}_0,\VEC{v}_0,0)$
fulfill a first law
\begin{equation}
\Delta E[\Traj{\VEC{x}},\Traj{\VEC{v}}] = Q[\Traj{\VEC{x}},\Traj{\VEC{v}}] + W[\Traj{\VEC{x}},\Traj{\VEC{v}}]
\, ,
\end{equation}
in correspondence to \eref{eq:1stLaw}.
A generalized definition of entropy production on the level of the GLE is
far less obvious; we thus discuss entropy production separately in Section \ref{sec:EP}.

The crucial \emph{mathematical} difference between the functionals
\eref{eq:QGLE}, \eref{eq:WGLE}
along the trajectory
$(\Traj{\VEC{x}},\Traj{\VEC{v}})=\{(\VEC{x}_t,\VEC{v}_t)\}_{t=0}^\tf$
generated by \eref{eq:GLE}
and those functionals in
\eref{eq:Q}, \eref{eq:W} is that now
the stochastic processes $(\VEC{x}_t,\VEC{v}_t)$
are sufficiently regular such that the integrals \eref{eq:QGLE}, \eref{eq:WGLE}
are \emph{independent of the discretization} chosen to evaluate them.
In other words, we do not need to prescribe any discretization scheme to render the
quantities $Q[\Traj{\VEC{x}},\Traj{\VEC{v}}]$ or $W[\Traj{\VEC{x}},\Traj{\VEC{v}}]$
uniquely defined, and there is no
ambiguity in their definition related to the interpretation of stochastic integrals.
Hence, we take these unique representations of heat and work
as our starting point to perform the same asymptotic analysis
which turns the GLE \eref{eq:GLE} into the overdamped Langevin equation \eref{eq:overdampedLE}.
As the main result we will 
obtain the asymptotic limits of the functionals \eref{eq:QGLE},
\eref{eq:WGLE} and ``automatically'' get delivered the correct discretization.

In order to streamline the computation of the asymptotic limits
and, at the same time, allow for similar functionals
to be analyzed in the same way, we subsume the heat and work
functionals under the general form
\begin{equation}
\label{eq:F}
\F[\Traj{\VEC{x}},\Traj{\VEC{v}}]
= \int_0^\tf r(\VEC{x}_t,\VEC{v}_t,t)\,\d t 
+ \int_0^\tf \VEC{q}(\VEC{x}_t,t) \cdot \d\VEC{x}_t
\, ,
\end{equation}
where $r(\VEC{x},\VEC{v},t)$ and $\VEC{q}(\VEC{x},t)$
are arbitrary (but ``well-behaved'') functions of the indicated arguments,
the latter vector-valued. The allocations
necessary to obtain \eref{eq:QGLE} and \eref{eq:WGLE} as special
cases of \eref{eq:F} are rather immediate.
For the heat $Q[\Traj{\VEC{x}},\Traj{\VEC{v}}]$ we have
$r(\VEC{x},\VEC{v},t)\equiv 0$ and
$\VEC{q}(\VEC{x},t)=-\VEC{f}(\VEC{x},t)$ and for the work $W[\Traj{\VEC{x}},\Traj{\VEC{v}}]$:
$r(\VEC{x},\VEC{v},t)\equiv \frac{\partial U(\VEC{x},t)}{\partial t} $ and
$\VEC{q}(\VEC{x},t)=-\fnc(\VEC{x},t)$.

\section{Formulation of the mathematical approach}
\label{sec:formulation}
The main mathematical difficulty in handling the GLE \eref{eq:GLE}
comes from the fact that it is a stochastic integro-differential equation driven by
a non-Markovian noise.
However, by extending its phase space and introducing,
in addition to the velocity degrees of freedom
\numparts
\begin{equation}
\label{eq:vdef}
\VEC{v}_t = \dot{\VEC{x}}_t
\, ,
\end{equation}
the auxiliary variables
\begin{equation}
\label{eq:ydef}
\VEC{y}_t = \int_0^t \d s \, {\kappa}(t-s) \TENSOR{h}(\VEC{x}_s)\dot{\VEC{x}}_s
\, ,
\end{equation}
\endnumparts
we can transform \eref{eq:GLE} to a system of Markovian
SDEs \cite{Lim2018}.
We here write this Markovian system in the differential form:
\numparts
\begin{eqnarray}
\d{\VEC{x}}_t
	& = & \VEC{v}_t \d t
\label{eq:dx}
\, , \\
\tau_v \d{\VEC{v}}_t
	& = & -\TENSOR{g}(\VEC{x}_t)\VEC{y}_t \d t + \frac{1}{\g}\VEC{f}(\VEC{x}_t,t) \d t
		+ \sqrt{2\kB T(\VEC{x}_t,t)/\g}\,\sigma(\VEC{x}_t)\VEC{\eta}_t \d t
\label{eq:dv}
\, , \\
\tb \d{\VEC{y}}_t
	& = & -\VEC{y}_t \d t + \TENSOR{h}(\VEC{x}_t)\VEC{v}_t \d t
\label{eq:dy}
\, , \\[1ex]
\tb \d{\VEC{\eta}}_t
	& = & -\VEC{\eta}_t \d t + \d\VEC{W}_t
\label{eq:deta}
\, .
\end{eqnarray}
\endnumparts
\changed{
The first equation \eref{eq:dx} is a transcription of \eref{eq:vdef}.
In the last equation \eref{eq:deta},
$\VEC{W}_t$ is a $d$-dimensional Wiener process,
whose (formal) derivative in time is the white noise $\VEC{\xi}_t$ from \eref{eq:etadot},
i.e.\ \eref{eq:deta} is a representation equivalent to \eref{eq:etadot}
of the dynamical equation generating the Ornstein-Uhlenbeck process $\VEC{\eta}_t$.
The third equation \eref{eq:dy} results from differentiating \eref{eq:ydef} with
respect to time and using the specific form \eref{eq:kappa} of $\kappa(t)$.
Note that also the more general case of a memory kernel $\kappa(t)$ which consists of a sum
of different exponentials can be recast into a Markovian system by introducing a
corresponding number of auxiliary variables like $\VEC{y}_t$ \cite{villamaina2009fluctuation},
and likewise one would
introduce a suitable number of $\VEC{\eta}_t$-like variables to represent a
combination of different Ornstein-Uhlenbeck processes driving the system of interest.
Finally, equation \eref{eq:dv} corresponds to the original GLE
\eref{eq:GLE} after all the auxiliary variables have been plugged in;
moreover, we use the abbreviation $\tau_v=m/\g$ for the characteristic equilibration
time of the velocity degrees of freedom.
}
These ``equations of motion'' \changed{\eref{eq:dv}-\eref{eq:deta}},
generating the Markov process
$(\VEC{x}_t,\VEC{v}_t,\VEC{y}_t,\VEC{\eta}_t)$ together with the SDE for the functional \eref{eq:F},
\begin{eqnarray}
\d\F_t & = &
r(\VEC{x}_t,\VEC{v}_t,t)\,\d t 
+ \VEC{q}(\VEC{x}_t,t)\cdot\d\VEC{x}_t
\nonumber \\
& = &
r(\VEC{x}_t,\VEC{v}_t,t)\,\changed{\d t}
+ \VEC{q}(\VEC{x}_t,t)\cdot\VEC{v}_t\,\d t
\label{eq:dF}
\, ,
\end{eqnarray}
constitute the basic set of equations for our asymptotic analysis.
In the second line of Eq.~\eref{eq:dF} we have introduced
$\d{\VEC{x}}_t=\VEC{v}_t \d t$ from \eref{eq:dx} in order to
turn the $\d\VEC{x}_t$ differentials into $\d t$ differentials as this will simplify
the analysis. We should nevertheless keep in mind that the $\VEC{q}(\VEC{x},t)$
term originated from a spatial differential, so that we might expect to get
back, from the asymptotic analysis, a contribution of the form $\VEC{q}\d\VEC{x}_t$
with some specific interpretation rule of the product.

We are interested in two different types of limiting procedures,
namely the white-noise limit and the small-mass limit.
In these asymptotic limits the time scales $\tb$ and $\tau_v$,
appearing in \eref{eq:dx}-\eref{eq:deta},
are ``much faster'' than the characteristic time scale $\tau_x$
of the diffusive motion in $\VEC{x}$.
Accordingly, we want to calculate the asymptotic equations resulting from
\eref{eq:dx}-\eref{eq:deta} and \eref{eq:dF} in the limits
$\tb/\tau_x \to 0$ and $\tau_v/\tau_x \to 0$.
The main idea is to exploit this scale separation by applying
a systematic multiscale or homogenization method \cite{pavliotis2008multiscale,bocquet1997high}
to the Fokker-Planck equation for the full density comprising fast
and slow variables, i.e.\ to
the Fokker-Planck equation for the density $\rho(t;\VEC{x},\VEC{v},\VEC{y},\VEC{\eta},\F)$
which is equivalent to the Langevin equations \eref{eq:dx}-\eref{eq:deta} and \eref{eq:dF}.
The result is an ``effective'' Fokker-Planck equation for the homogenized density of
the slow variables alone.

It is, however, not obvious how the above two limits
relate to each other, i.e.~if they are joint limits or if there is
a specific order in which they have to be taken.
To answer this question we have to consider the specific physical system we
intend to model with \eref{eq:dx}-\eref{eq:deta}.
Having in mind a colloidal particle (the system) suspended in aqueous solution
at room temperature (the thermal bath), the time scale $\tb$ of the bath-intrinsic correlations
is related to the molecular collision time of the water molecules, which is of order $\tb \approx \unit[10^{-13}]{s}$.
The velocity relaxation time is given by the ratio of inertia and friction effects and
amounts to $\tau_v \approx \unit[0.5 \times 10^{-8}]{s}$ for a colloid of $\unit[0.1]{\mu m}$ radius,
while the colloid's diffusive time scale is of the order $\tau_x \approx \unit[0.5 \times 10^{-2}]{s}$
\cite{philipse2018brownian}.
In other words, all three time scales are in fact well-separated,
$\tb \ll \tau_v \ll \tau_x$, and are governed by different physical processes.
Hence, we should perform the limits separately in two steps:
\begin{enumerate}
\item[1)]
$\tb/\tau_x \to 0$, where $\tau_v/\tau_x$ remains constant and finite,
\item[2)]
$\tau_v/\tau_x \to 0$.
\end{enumerate}
We summarize the results for the first limit (white-noise or Markovian limit)
in Section \ref{sec:WNlimit}. 
The second limit (small-mass limit) is treated in Section \ref{sec:SMlimit}.
For both cases, the technical details are given in \ref{app:HFPE}.

In order to perform the homogenization procedure we need to know
how the various terms in \eref{eq:dx}-\eref{eq:deta} and \eref{eq:dF} scale
with the relevant time- and length scales.
We therefore rewrite \eref{eq:dx}-\eref{eq:deta} and \eref{eq:dF} by
introducing dimensionless quantities which are all of order one,
so that the dependencies on scales show up as
dimensionless (pre-) factors which we expect to contain $\tb$, $\tau_v$ and $\tau_x$.
Marking dimensionless quantities by a tilde,
we define
\numparts
\begin{eqnarray}
t = \tau_x \tilde{t}
\label{eq:tscale}
\, , \\
\VEC{x} = L \tilde{\VEC{x}}
\label{eq:xscale}
\, , \\
\VEC{v} = \sqrt{\frac{\kB \bar{T}}{m}} \tilde{\VEC{v}} = \frac{L}{\sqrt{\tau_x \tau_v}} \tilde{\VEC{v}}
\label{eq:vscale}
\, , \\
\VEC{y} = \frac{L}{\sqrt{\tau_x \tb}} \tilde{\VEC{y}}
\label{eq:yscale}
\, , \\
\VEC{\eta} = \frac{1}{\sqrt{\tb}} \tilde{\VEC{\eta}}
\label{eq:etascale}
\, , \\
\VEC{f} = \frac{\kB \bar{T}}{L} \tilde{\VEC{f}} \quad\Rightarrow\quad \frac{\VEC{f}}{\g} = \frac{L}{\tau_x} \tilde{\VEC{f}}
\label{eq:fscale}
\, .
\end{eqnarray}
\endnumparts
We measure the variables of interest $t$ and $\VEC{x}$ on the slowest scale of the system,
namely the diffusive time scale $\tau_x$ and an associated ``large'' length-scale $L$, such that
\begin{equation}
\tau_x = \frac{L^2}{\bar{D}} = \frac{\g L^2}{\kB \bar{T}}
\, ,
\end{equation}
where $\bar{T}$ quantifies the typical temperature of the bath according to
$T(\VEC{x},t)=\bar{T}\Theta(\VEC{x},t)$,
with a dimensionless temperature profile $\Theta(\VEC{x},t)$.
The velocity degrees of freedom are of the order of the thermal velocity, since
they equilibrate on the scale $\tau_v$ much faster than $\tau_x$.
The typical magnitude of the auxiliary variable $\VEC{\eta}$ is obtained from its
definition \eref{eq:ydef} by rewriting the integral in a form which is independent
of all scales involved.
The order of magnitude for $\VEC{\eta}$ follows directly from its correlations
\eref{eq:etaietaj}.
We assume the external force $\VEC{f}$ to vary on slow scales only and to correspond to
energies comparable to the typical thermal energy of the bath.
The tensors $\TENSOR{g}$, $\TENSOR{h}$, $\TENSOR{\sigma}$ are already dimensionless,
because of the prefactor $\g$ in \eref{eq:GLE}.

Plugging \eref{eq:tscale}-\eref{eq:fscale} into the GLE \eref{eq:dx}-\eref{eq:deta},
we find (omitting arguments):
\numparts
\begin{eqnarray}
\d\tilde{\VEC{x}}_{\tilde{t}}
	& = & \sqrt{\frac{\tau_x}{\tau_v}}\tilde{\VEC{v}}_{\tilde{t}} \,\d \tilde{t}
\label{eq:tildedx}
\, , \\
\d\tilde{\VEC{v}}_{\tilde t}
	& = & -\frac{\tau_x}{\sqrt{\tb \tau_v}}\TENSOR{g}\tilde{\VEC{y}}_{\tilde{t}} \,\d \tilde{t}
		+ \sqrt{\frac{\tau_x}{\tau_v}}\tilde{\VEC{f}} \d \tilde{t}
		+ \sqrt{2 \Theta}\,\frac{\tau_x}{\sqrt{\tb \tau_v}}\sigma \tilde{\VEC{\eta}}_{\tilde{t}} \,\d \tilde{t}
\label{eq:tildedv}
\, , \\
\d\tilde{\VEC{y}}_{\tilde{t}}
	& = & -\frac{\tau_x}{\tb}\tilde{\VEC{y}}_{\tilde{t}} \,\d \tilde{t}
		+ \frac{\tau_x}{\sqrt{\tb \tau_v}}\TENSOR{h}\tilde{\VEC{v}}_{\tilde{t}} \,\d \tilde{t}
\label{eq:tildedy}
\, , \\[1ex]
\d\tilde{\VEC{\eta}}_{\tilde{t}}
	& = & -\frac{\tau_x}{\tb}\tilde{\VEC{\eta}}_{\tilde{t}} \,\d\tilde{t}
		+ \sqrt{\frac{\tau_x}{\tb}} \,\d\tilde{\VEC{W}}_{\tilde{t}}
\label{eq:tildedeta}
\, .
\end{eqnarray}
\endnumparts
Similarly, we find the dimensionless form of the general functional \eref{eq:dF},
which includes the dimensionless representations of heat and work
from \eref{eq:QGLE}, \eref{eq:WGLE}, to be
\begin{equation}
\label{eq:dFtilde}
\d\tilde{\F}_{\tilde t} =
	\tilde{r}\,\d\tilde{t} 
	+ \sqrt{\frac{\tau_x}{\tau_v}}\, \tilde{\VEC{q}}\cdot\tilde{\VEC{v}}_{\tilde{t}}\,\d\tilde{t}
\, ,
\end{equation}
with the dimensionless counterparts
$\tilde{r}=\tilde{r}(\tilde{\VEC{x}}_{\tilde{t}},\tilde{\VEC{v}}_{\tilde{t}},\tilde{t})$,
$\tilde{\VEC{q}}=\tilde{\VEC{q}}(\tilde{\VEC{x}}_{\tilde{t}},\tilde{t})$
of the functions
$r(\VEC{x}_t,\VEC{v}_t,t)$, $\VEC{q}(\VEC{x}_t,t)$
from \eref{eq:dF}.

The Fokker-Planck equation associated to the SDEs \eref{eq:tildedx}-\eref{eq:dFtilde} is:
\numparts
\begin{equation}
\label{eq:FPE}
\frac{\partial\rho}{\partial\tilde{t}} = -\tilde{\nabla}\cdot\tilde{\VEC{J}}
\, ,
\end{equation}
for the density $\rho=\rho(\tilde{t};\tilde{\VEC{x}},\tilde{\VEC{v}},\tilde{\VEC{y}},\tilde{\VEC{\eta}},\tilde{\F})$
with
\begin{equation}
\tilde{\nabla} =
\left(
\begin{array}{c}
\nabla_{\tilde{\VEC{x}}}
\\
\nabla_{\tilde{\VEC{v}}}
\\
\nabla_{\tilde{\VEC{y}}}
\\
\nabla_{\tilde{\VEC{\eta}}}
\\
\partial/\partial\tilde{\F}
\end{array}
\right)
\end{equation}
and
\begin{equation}
\label{eq:J}
\tilde{\VEC{J}} = 
\left(
\begin{array}{c}
\sqrt{\frac{\tau_x}{\tau_v}}\tilde{\VEC{v}}
\\[0.5ex]
-\frac{\tau_x}{\sqrt{\tb \tau_v}}\TENSOR{g}\tilde{\VEC{y}}
	+ \sqrt{\frac{\tau_x}{\tau_v}}\tilde{\VEC{f}}
	+ \sqrt{2 \Theta}\,\frac{\tau_x}{\sqrt{\tb \tau_v}}\sigma \tilde{\VEC{\eta}}
\\[0.5ex]
-\frac{\tau_x}{\tb}\tilde{\VEC{y}}
	+ \frac{\tau_x}{\sqrt{\tb \tau_v}}\TENSOR{h}\tilde{\VEC{v}}
\\[0.5ex]
-\frac{\tau_x}{\tb}\tilde{\VEC{\eta}}
		- \frac{\tau_x}{2\tb} \nabla_{\tilde{\VEC{\eta}}}
\\[0.5ex]
\tilde{r} + \sqrt{\frac{\tau_x}{\tau_v}}\, \tilde{\VEC{q}}\cdot\tilde{\VEC{v}}
\end{array}
\right)\rho
\, .
\end{equation}
\endnumparts
In the following, we apply the two-step limiting procedure as described
above to this Fokker-Planck equation. The technical details are given in
\ref{app:HFPE}.

\section{The homogenized Fokker-Planck and Langevin equation}
\label{sec:limits}
\subsection{The white-noise limit}
\label{sec:WNlimit}
In the white-noise limit we have $\tb/\tau_x \to 0$ while $\tau_v/\tau_x$
stays finite. Defining the small parameter
\begin{equation}
\epsilon = \sqrt{\frac{\tb}{\tau_x}}
\end{equation}
we can write the Fokker-Planck equation \eref{eq:FPE}-\eref{eq:J}
in the form
\begin{equation}
\label{eq:FPEeps}
\frac{\partial\rho}{\partial\tilde{t}} =
-\left(
	\OP{L}_2^\dagger + \frac{1}{\epsilon}\OP{L}_1^\dagger + \frac{1}{\epsilon^2}\OP{L}_0^\dagger
\right)\rho
\end{equation}
\numparts
with the operators
\begin{eqnarray}
\OP{L}_2^\dagger =
\sqrt{\frac{\tau_x}{\tau_v}}\, \tilde{v}_i \frac{\partial}{\partial\tilde{x}_i}
+ \sqrt{\frac{\tau_x}{\tau_v}}\, \tilde{f}_i \frac{\partial}{\partial\tilde{v}_i}
+ \left( \tilde{r} + \sqrt{\frac{\tau_x}{\tau_v}}\, \tilde{\VEC{q}}\cdot\tilde{\VEC{v}} \right) \frac{\partial}{\partial\tilde{\F}}
\label{eq:WNL0}
\, , \\
\OP{L}_1^\dagger =
-\sqrt{\frac{\tau_x}{\tau_v}}\, g_{ij} \tilde{y}_j \frac{\partial}{\partial\tilde{v}_i}
+ \sqrt{\frac{\tau_x}{\tau_v}}\, \sqrt{2\Theta}\, \sigma_{ij}\tilde{\eta}_j \frac{\partial}{\partial\tilde{v}_i}
+ \sqrt{\frac{\tau_x}{\tau_v}}\, h_{ij} \tilde{v}_j \frac{\partial}{\partial\tilde{y}_i}
\label{eq:WNL1}
\, , \\
\OP{L}_0^\dagger =
-\frac{\partial}{\partial\tilde{y}_i}\tilde{y}_i
- \frac{\partial}{\partial\tilde{\eta}_i}\tilde{\eta}_i
- \frac{1}{2}\frac{\partial^2}{\partial\tilde{\eta}_i^2}
\label{eq:WNL2}
\, ,
\end{eqnarray}
\endnumparts
where we use Einstein's summation convention over repeated indices.
Performing the multi-scale procedure outlined in \ref{app:HFPE}
we find the ``effective'' Fokker-Planck equation
\begin{eqnarray}
\frac{\partial\rho}{\partial\tilde{t}} =
-\left[
	\sqrt{\frac{\tau_x}{\tau_v}}\, \tilde{v}_i \frac{\partial}{\partial\tilde{x}_i}
	- \frac{\tau_x}{\tau_v}\, g_{ij}h_{jk} \frac{\partial}{\partial\tilde{v}_i} \tilde{v}_k
	+ \sqrt{\frac{\tau_x}{\tau_v}}\, \tilde{f}_i \frac{\partial}{\partial\tilde{v}_i}
\right.
\nonumber \\
\left. \qquad\qquad \mbox{}
	-\frac{\tau_x}{\tau_v}\, \Theta\, \sigma_{ij}(\sigma\TRANS)_{jk}
		\frac{\partial}{\partial\tilde{v}_i}\frac{\partial}{\partial\tilde{v}_k}
	+ \left( \tilde{r} + \sqrt{\frac{\tau_x}{\tau_v}}\, \tilde{\VEC{q}}\cdot\tilde{\VEC{v}} \right) \frac{\partial}{\partial\tilde{\F}}
\right] \rho
\label{eq:underdampedFPE}
\end{eqnarray}
which governs the dynamics of the reduced density
$\rho=\rho(\tilde{t},\tilde{\VEC{x}},\tilde{\VEC{v}},\tilde{\F})$ for the
slow variables $(\tilde{\VEC{x}},\tilde{\VEC{v}})$ in the asymptotic
limit $\epsilon \to 0$.
It is equivalent to the Langevin equations
\numparts
\begin{eqnarray}
\d\tilde{\VEC{x}}_{\tilde{t}}
	& = & \sqrt{\frac{\tau_x}{\tau_v}}\tilde{\VEC{v}}_{\tilde{t}} \,\d \tilde{t}
\, , \\
\d\tilde{\VEC{v}}_{\tilde{t}}
	& = & -\frac{\tau_x}{\tau_v} \TENSOR{g}\TENSOR{h}\tilde{\VEC{v}}_{\tilde{t}} \,\d \tilde{t}
			+ \sqrt{\frac{\tau_x}{\tau_v}} \tilde{\VEC{f}} \,\d \tilde{t}
			+ \sqrt{\frac{\tau_x}{\tau_v}} \, \sqrt{2\Theta}\, \TENSOR{\sigma} \,\d\tilde{\VEC{W}}_{\tilde{t}}
\, , \\
\d\tilde{\F}_{\tilde t}
	& = & \left( \tilde{r} + \sqrt{\frac{\tau_x}{\tau_v}}\, \tilde{\VEC{q}}\cdot\tilde{\VEC{v}}_{\tilde{t}} \right) \,\d \tilde{t}
\, .
\end{eqnarray}
\endnumparts
For $\TENSOR{g}=\TENSOR{h}\TRANS=\TENSOR{\sigma}=\TENSOR{\gamma}^{1/2}$
these are exactly the standard Langevin-Kramers equations (see e.g. \cite{mazo2002brownian})
describing underdamped Brownian motion, together with the same equation for the
functional as we had it in the very beginning.
Since none of the functions appearing in the Langevin-Kramers equation or in the
functional depend on the (auxiliary) variables $(\VEC{y},\VEC{\eta})$, which have
been ``integrated out'' during the homogenization procedure, the Markovian limit
$\tb/\tau_x \to 0$ is basically trivial in the current context.
As we will see in the next Section, however, the small-mass limit $\tau_v/\tau_x \to 0$
is much more interesting, in particular in our case in which
$\TENSOR{g}$, $\TENSOR{h}$, $\TENSOR{\sigma}$, and $\Theta$ depend on the
particle position.

\subsection{The small-mass limit}
\label{sec:SMlimit}
For the small-mass limit the small parameter is
\begin{equation}
\epsilon = \sqrt{\frac{\tau_v}{\tau_x}}
\, .
\end{equation}
Using this definition, we can bring the Fokker-Planck equation \eref{eq:underdampedFPE}
for the underdamped motion into the form \eref{eq:FPEeps} with the operators
\numparts
\begin{eqnarray}
\OP{L}_2^\dagger =
\tilde{r} \frac{\partial}{\partial\tilde{\F}}
\label{eq:SML0}
\, , \\
\OP{L}_1^\dagger =
\tilde{v}_i \frac{\partial}{\partial\tilde{x}_i}
+ \tilde{f}_i \frac{\partial}{\partial\tilde{v}_i}
+ \tilde{\VEC{q}}\cdot\tilde{\VEC{v}} \frac{\partial}{\partial\tilde{\F}}
\label{eq:SML1}
\, , \\
\OP{L}_0^\dagger =
-(\TENSOR{g}\TENSOR{h})_{ij} \frac{\partial}{\partial\tilde{v}_i} \tilde{v}_j
- \Theta \, (\sigma\sigma\TRANS)_{ij} \frac{\partial}{\partial\tilde{v}_i}\frac{\partial}{\partial\tilde{v}_j}
\label{eq:SML2}
\, .
\end{eqnarray}
\endnumparts
The homogenization procedure from \ref{app:HFPE} then yields the
``effective'' Fokker-Planck equation
\begin{eqnarray}
\frac{\partial\rho}{\partial\tilde{t}} =
-\left[
	\frac{\partial}{\partial\tilde{x}_i} [(\TENSOR{g}\TENSOR{h})^{-1}]_{ij} \tilde{f}_j
	- \frac{\partial}{\partial\tilde{x}_i} [(\TENSOR{g}\TENSOR{h})^{-1}]_{ij} \frac{\partial}{\partial\tilde{x}_k} \Sigma_{jk}\Theta
\right.
\nonumber \\ \qquad\qquad \mbox{}
	+ \tilde{r} \frac{\partial}{\partial\tilde{\F}}
	+ \tilde{q}_i [(\TENSOR{g}\TENSOR{h})^{-1}]_{ij} \tilde{f}_j \frac{\partial}{\partial\tilde{\F}}
	+ \Theta \frac{\partial \tilde{q}_i [(\TENSOR{g}\TENSOR{h})^{-1}]_{ij}}{\partial\tilde{x}_k}
		\Sigma_{jk} \frac{\partial}{\partial\tilde{\F}}
\nonumber \\ \qquad\qquad \mbox{}
	- \frac{\partial}{\partial\tilde{x}_i} \Theta [(\TENSOR{g}\TENSOR{h})^{-1}]_{ij} \Sigma_{jk} \tilde{q}_k
		\frac{\partial}{\partial\tilde{\F}}
	- \frac{\partial}{\partial\tilde{x}_k} \Theta \tilde{q}_i [(\TENSOR{g}\TENSOR{h})^{-1}]_{ij} \Sigma_{jk}
		\frac{\partial}{\partial\tilde{\F}}
\nonumber \\ \qquad\qquad\qquad\qquad\qquad\qquad\qquad\quad \mbox{}
\left.
	- \tilde{q}_i [(\TENSOR{g}\TENSOR{h})^{-1}]_{ij} \Sigma_{jk} \tilde{q}_k \Theta \frac{\partial^2}{\partial\tilde{\F}^2}
\right] \rho
\label{eq:overdampedFPE}
\, ,
\end{eqnarray}
where it is understood that a derivative with a function being included in
the numerator of the derivative's fraction acts only on that function,
but not on anything else that appears further to the right.
The symmetric $d \times d$ matrix $\TENSOR{\Sigma}$ is proportional to the
covariance of the Gaussian distribution, which is the stationary distribution of
the ``fast process'' \eref{eq:SML2} (at a given $\tilde{\VEC{x}}$) 
and solves the Lyapunov equation
\begin{equation}
\label{eq:Sigma}
(\TENSOR{g}\TENSOR{h})\TENSOR{\Sigma} + \TENSOR{\Sigma}(\TENSOR{g}\TENSOR{h})\TRANS = 2\TENSOR{\sigma}\TENSOR{\sigma}\TRANS
\, .
\end{equation}
In the above, we have implicitly assumed that $\TENSOR{g}\TENSOR{h}$ is positive definite
(and therefore symmetric and invertible) so that the Lyapunov equation has a unique solution,
which can be written as \cite{bellman1997introduction}
\begin{equation} 
\TENSOR{\Sigma} =
2 \int_0^\infty e^{- \TENSOR{g}\TENSOR{h} z} \TENSOR{\sigma} \TENSOR{\sigma}\TRANS
	e^{- (\TENSOR{g}\TENSOR{h})\TRANS z} \,\d z. 
\end{equation}
Note that this assumption is satisfied for the physically relevant case in which $\TENSOR{g}\TENSOR{h}$
is a friction tensor, since it then must be positive definite. If, in addition, $\TENSOR{g}\TENSOR{h}$ commutes with $\TENSOR{\sigma}\TENSOR{\sigma}\TRANS$
(i.e., if the fast process governed by \eref{eq:SML2} satisfies detailed balance),
then we have $\TENSOR{\Sigma} = (\TENSOR{g}\TENSOR{h})^{-1} \TENSOR{\sigma}\TENSOR{\sigma}\TRANS$. It is easy to see that the
fluctuation-dissipation relation ($\TENSOR{g} = \TENSOR{h}\TRANS = \TENSOR{\sigma}$)
implies detailed balance but the converse is generally not true.

The Fokker-Planck equation \eref{eq:overdampedFPE} describes the time evolution of
the density $\rho=\rho(\tilde{t},\tilde{\VEC{x}},\tilde{\F})$
for the slow variables $\tilde{\VEC{x}}$ and $\tilde{\F}$.
The first line represents the $\tilde{\VEC{x}}$-dynamics.
The space-dependent terms involving the strength of thermal fluctuations, i.e., the temperature $\Theta$ and  $\Sigma$
appear in It\^o position, i.e.~to the right of both derivatives in the
diffusion term, meaning that with a constant friction the correct interpretation of the multiplicative noise would be the It\^o one (start-point rule). The space-dependent (inverse) friction coefficient $(\TENSOR{g}\TENSOR{h})^{-1}$
appears in between these two derivatives and thus as an anti-It\^o or H\"anggi-Klimontovich term. 
Such splitting of friction and temperature into anti-It\^o and It\^o contributions is
well known \cite{sancho1982adiabatic,jayannavar1995macroscopic,yang2013brownian,hottovy2012noise,hottovy2015smoluchowski}
for a locally equilibrium bath with the fluctuation-dissipation relation being valid at each position $\VEC{x}$,
for which $\TENSOR{h}\TENSOR{g}=\TENSOR{\sigma}\TENSOR{\sigma}\TRANS=\tilde{\TENSOR{\gamma}}$ are identical
to the (dimensionless) friction tensor $\tilde{\TENSOR{\gamma}}$, and for which $\TENSOR{\Sigma}=\TENSOR{1}$ (see Eq.~\eref{eq:Sigma}).
The second to fourth line in \eref{eq:overdampedFPE} characterize the dynamics of the functional $\tilde{\F}$.
If $(\TENSOR{g}\TENSOR{h})^{-1}\Sigma$ is symmetric, which is guaranteed if the fast process obeys detailed balance, 
 these terms are consistent with a contribution $\tilde{\VEC{q}}\circ\d\tilde{\VEC{x}}$
to the drift of the functional $\tilde{\F}$, interpreted in the Stratonovich sense,
see \ref{app:joint}.
In fact, the third term in the second line contains both, the spurious drift
$\Theta \tilde{q}_i \frac{\partial [(\TENSOR{g}\TENSOR{h})^{-1}]_{ij}}{\partial\tilde{x}_k}\Sigma_{jk}$
which appears in $\d\tilde{\VEC{x}}$ when writing its thermal noise term in It\^o-form,
as well as the correction 
$\Theta \frac{\partial\tilde{q}_i}{\partial\tilde{x}_k}[(\TENSOR{g}\TENSOR{h})^{-1}]_{ij}\Sigma_{jk}$
that changes $\tilde{\VEC{q}}\circ\d\tilde{\VEC{x}}$ from Stratonovich form to
$\tilde{\VEC{q}}\cdot\d\tilde{\VEC{x}}$ in It\^o form.

The set of Langevin equations equivalent to \eref{eq:overdampedFPE} therefore
reads
\numparts
\begin{eqnarray}
\d\VEC{x}_t =
	\frac{1}{\g}(\TENSOR{g}(\VEC{x}_t)\TENSOR{h}(\VEC{x}_t))^{-1} \VEC{f}(\VEC{x}_t,t) \,\d t
  + 
  \VEC{s}(\VEC{x}_t,t) \,\d t
\nonumber \\ \qquad\qquad \mbox{}
	+ \sqrt{2\kB T(\VEC{x}_t,t)/\g}\, [(\TENSOR{g}(\VEC{x}_t)\TENSOR{h}(\VEC{x}_t))^{-1}\TENSOR{\Sigma}(\VEC{x}_t)]^{1/2}
	\,\d\VEC{W}_t
\label{eq:dxWNSM}
\, , \\[1ex]
\d\F_t = r(\VEC{x}_t,t) \,\d t + \VEC{q}(\VEC{x}_t,t)\circ\d\VEC{x}_t
\label{eq:dFWNSM}
\, ,
\end{eqnarray}
where we switched back to dimensionful quantities.
The term $\VEC{s}(\VEC{x},t)$ represents a noise-induced drift
corresponding to an It\^o-interpreted multiplicative noise in $\d\VEC{x}_t$.
Its $i$-th component is given by
\begin{equation}
\label{eq:siWNSM} 
s_i(\VEC{x},t) =
	\frac{\kB T(\VEC{x},t)}{\g}
	\frac{\partial[(\TENSOR{g}(\VEC{x})\TENSOR{h}(\VEC{x}))^{-1}]_{ij}}{\partial x_k} \Sigma_{jk}
\end{equation}
\endnumparts
with $\TENSOR{\Sigma}$ being the solution of the Lyapunov equation \eref{eq:Sigma}.
These equations are the main result of the present contribution.
Our systematic and general approach considering the white-noise and small-mass limit
for the dynamics and for thermodynamical functionals simultaneously
also recovers some of the results obtained in a series of related works,
both for the dynamics
\cite{sancho1982adiabatic,hanggi1994colored,jayannavar1995macroscopic,yang2013brownian,kupferman2004ito,hottovy2012noise,hottovy2015smoluchowski,pavliotis2008multiscale,stolovitzky1998non,moon2014interpretation,wong1965convergence,bo2013white},
and the stochastic thermodynamics for entropy \cite{celani2012anomalous,spinney2012entropy,kawaguchi2013fluctuation,bo2014entropy,nakayama2015invariance,ford2015stochastic,marino2016entropy,ge2018anomalous,miyazaki2018entropy,birrell2018entropy},
heat \cite{bo2013entropic,ge2014time}, work \cite{PhysRevE.98.052105} or general functionals \cite{bo2017multiple}.
The most important observation is that a functional of the underdamped dynamics
$(\VEC{x},\VEC{v})$
with a differential of the form $\VEC{q}(\VEC{x}_t,t)\cdot\d\VEC{x}_t=\VEC{q}(\VEC{x}_t,t)\cdot\VEC{v_t}\d t$
($\VEC{q}$ being independent of $\VEC{v}$) in the small-mass limit always reduces to
$\VEC{q}(\VEC{x}_t,t)\circ\d\VEC{x}_t$ with a Stratonovich
product. This behavior occurs completely independent of spatial dependencies of the friction
tensors or the temperature field, but rests on two important conditions: i) the processes which generate the colored noise
to be equilibrium ones, ii) in the small-mass limit the velocity degrees of freedom reach an equilibrium distribution with the local temperature (this is always guaranteed if the system obeys  the
fluctuation-dissipation relation $\TENSOR{g} = \TENSOR{h}\TRANS = \TENSOR{\sigma}$) \cite{pavliotis2008multiscale,bo2013white}. 

As a specific example of \eref{eq:dxWNSM}-\eref{eq:siWNSM},
we consider the heat and work functionals \eref{eq:QGLE} and \eref{eq:WGLE}
for a Langevin equation, which fulfills the fluctuation-dissipation relation,
$\g\TENSOR{h}\TENSOR{g}=\g\TENSOR{\sigma}\TENSOR{\sigma}\TRANS=\g\tilde{\TENSOR{\gamma}}=\TENSOR{\gamma}$:
\numparts
\begin{eqnarray}
\d\VEC{x}_t =
	\TENSOR{\gamma}^{-1}(\VEC{x}_t) \VEC{f}(\VEC{x}_t,t) \,\d t
	+ \kB T(\VEC{x}_t,t) [\nabla_{\VEC{x}_t}\TENSOR{\gamma}^{-1}(\VEC{x}_t)]\,\d t
\nonumber \\ \qquad\qquad\qquad \mbox{}
	+ \sqrt{2\kB T(\VEC{x}_t,t)}\, \TENSOR{\gamma}^{-1/2}(\VEC{x}_t) \,\d\VEC{W}_t
\label{eq:overampedLEDB}
\, , \\[1ex]
\d Q_t =
	-\VEC{f}(\VEC{x}_t,t)\circ\d\VEC{x}_t
  = \Big( \nabla_{\VEC{x}_t} U(\VEC{x}_t,t) - \fnc(\VEC{x}_t,t) \Big)\circ\d\VEC{x}_t
\label{eq:overampedHeatDB}
\, , \\[1ex]
\d W_t= \frac{\partial U(\VEC{x}_t,t)}{\partial t}\,\mathrm{d}t + \fnc(\VEC{x}_t,t) \circ \d\VEC{x}_t
\, ,
\label{eq:overampedWorkDB}
\end{eqnarray}
\endnumparts
where the multiplicative noise in $\d\VEC{x}_t$ is an It\^o product.
The Stratonovich product in $\d Q_t$ and $\d W_t$ confirms our physically motivated arguments
from the Introduction, based on viewing heat, work etc.\ as transformations of the
dynamical variables \cite{sekimoto2010stochastic}.
The original definition of heat was based on the identification with the energy exchanges 
due to the frictional 
and fluctuating force 
exerted by the heat bath on the system \cite{sekimoto2010stochastic}. 
This interpretation is preserved also in the case of space-dependent friction
or temperature with the additional condition that any spurious (noise-induced)
drift term in the Langevin equation, which might
result from a certain representation of the multiplicative noise term, always has to
be interpreted to contribute to heat (and not to work).
This is testified by the fact that the limiting expression \eref{eq:overampedHeatDB} is identical to
\eref{eq:Q} and there is no correction to $\VEC{f}$ due to space-dependent friction
or temperature.
In other words, the specific interpretation of the multiplicative noise in 
the overdamped Langevin equation with space-dependent
friction or temperature 
is irrelevant for the identification of
the contributions of the various drift terms to heat and work,
a quite comforting finding from a physical perspective.

\section{Entropy production}
\label{sec:EP}
For the sake of completeness, we here discuss homogenization of entropy production.
It has been shown \cite{celani2012anomalous}
that, in presence of temperature gradients, the homogenized Fokker-Planck
equation for entropy production acquires non-trivial terms which describe a so-called ``entropic anomaly''
of the small-mass limit, see also
\cite{spinney2012entropy,kawaguchi2013fluctuation,bo2014entropy,marino2016entropy,ge2018anomalous,miyazaki2018entropy,bo2017multiple,bo2013entropic,ge2014time,birrell2018entropy}.
In the following, we briefly summarize the calculation and main results from
\cite{celani2012anomalous},
treating explicitly the case in which the friction tensor is not simply
proportional to the identity, see also \cite{marino2016entropy}.

In stochastic thermodynamics the entropy production (in the thermal environment) is
usually defined \cite{kurchan1998fluctuation,seifert2012stochastic,chetrite2008fluctuation}
as a measure of irreversibility via the log-ratio of probabilities
for observing a specific trajectory $(\Traj{\VEC{x}},\Traj{\VEC{v}})=\{(\VEC{x}_t,\VEC{v}_t)\}_{t=0}^{\tf}$
in forward time versus observing the same trajectory traced out backwards when advancing time.
For the overdamped Langevin equation,
the explicit expression \eref{eq:DeltaS} for the resulting functional is obtained
from path-integral techniques \cite{kurchan1998fluctuation,seifert2012stochastic,chetrite2008fluctuation}.
The designation \emph{entropy production in the thermal environment}
for this irreversibility measure originates from the central observation that
it is equivalent to the heat dissipated into the bath along the trajectory
divided by the bath temperature, see our discussion of \eref{eq:DeltaS}.

However, for the GLE \eref{eq:GLE} a direct calculation using path integrals
of path probabilities involving only $(\Traj{\VEC{x}},\Traj{\VEC{v}})$
(without explicit knowledge of the auxiliary variables $(\Traj{\VEC{y}},\Traj{\VEC{\eta}})$),
is much more challenging due to the GLE's non-Markovian character,
and has not been achieved yet to the best of our knowledge.
To analyze entropy production, we therefore start
from its well-known expression for the Langevin-Kramers equation,
\begin{eqnarray}
\Delta S^{\mathrm{env}}[\Traj{\VEC{x}},\Traj{\VEC{v}}]
& = & \int_0^\tf \frac{1}{T(\VEC{x}_t,t)} \Big( \VEC{f}(\VEC{x}_t,t) - m\dot{\VEC{v}}_t \Big)\circ\d\VEC{x}_t
\nonumber \\[1ex]
& = & \int_0^\tf \frac{1}{T(\VEC{x}_t,t)} \Big( \VEC{f}(\VEC{x}_t,t)\cdot\d\VEC{x}_t - m{\VEC{v}}_t\circ\d\VEC{v}_t \Big)
\, ,
\label{eq:DeltaSGLE}
\end{eqnarray}
and assume that this expression
is the asymptotic limit of a (unknown) ``generalized entropy production''
for vanishing noise correlation time $\tb/\tau_x \to 0$.

For analyzing the small-mass limit $\tau_v/\tau_x \to 0$,
it proves convenient to split off a boundary term,
because we then can rewrite the $\d\VEC{v}_t$ integral
in \eref{eq:DeltaSGLE} as a sum of $\d t$ and $\d\VEC{x}_t$ integrals \cite{celani2012anomalous,marino2016entropy}.
To do so, we first introduce
the Maxwell-Boltzmann velocity distribution at given position $\VEC{x}$ and time $t$,
\begin{equation}
\label{eq:MB}
w = \left( \frac{m}{2\pi \kB T(\VEC{x},t)} \right)^{d/2} \exp\left[ {-\frac{m\VEC{v}^2}{2\kB T(\VEC{x},t)}} \right]
\, .
\end{equation}
By observing that the total differential
$-\d\ln w = \d( \frac{m\VEC{v}^2}{2\kB T} + \frac{d}{2}\ln\frac{2\pi\kB T}{m} )$
indeed contains the desired term $\frac{m\VEC{v}_t}{\kB T}\cdot\d\VEC{v}_t$,
we can eliminate it from the integral in \eref{eq:DeltaSGLE} and find
\begin{eqnarray}
\Delta S^{\mathrm{env}}[\Traj{\VEC{x}},\Traj{\VEC{v}}] - \kB \ln\frac{w_{\mathrm{f}}}{w_0}
= \int_0^\tf \left[
	\frac{1}{T(\VEC{x}_t,t)} \VEC{f}(\VEC{x}_t,t)\cdot\d\VEC{x}_t
\right.
\nonumber \\ \qquad\quad \left. \mbox{}
+ \frac{d\,\kB T(\VEC{x}_t,t)-m\VEC{v}_t^2}{2T(\VEC{x}_t,t)^2}
	\left( \frac{\partial T(\VEC{x}_t,t)}{\partial t}\d t + \frac{\partial T(\VEC{x}_t,t)}{\partial\VEC{x}}\cdot\d\VEC{x}_t \right)
\right]
\label{eq:DeltaSGLEx}
\, ,
\end{eqnarray}
where $w_0$ and $w_{\mathrm{f}}$ denote the Maxwell-Boltzmann distribution
\eref{eq:MB} at initial and final point of the trajectory.

The corresponding dimensionless differential of the functional reads (omitting arguments)
\begin{equation}
\label{eq:dFEP}
\d\tilde{\F} =
\left[
	\sqrt{\frac{\tau_x}{\tau_v}}\frac{\tilde{\VEC{f}}\cdot\tilde{\VEC{v}}}{\Theta}
	+ \frac{d\Theta-\tilde{\VEC{v}}^2}{2\Theta^2}
	\left(
		\frac{\partial\Theta}{\partial\tilde{t}} + \sqrt{\frac{\tau_x}{\tau_v}} \nabla_{\tilde{\VEC{x}}}\Theta\cdot\tilde{\VEC{v}}
	\right)
\right] \d\tilde{t}
\, .
\end{equation}
Our goal is to perform the small mass asymptotic limit $\epsilon=\sqrt{\tau_v/\tau_x} \to 0$
of this functional
in conjunction with the Langevin-Kramers equation, exactly as we have done it in Section
\ref{sec:SMlimit} for heat- and work-like functionals.
We will here, however, focus on the case when  
$\TENSOR{h}\TENSOR{g}=\TENSOR{\sigma}\TENSOR{\sigma}\TRANS=\tilde{\TENSOR{\gamma}}$ is positive definite,
such that the fluctuation-dissipation relation is fulfilled
(otherwise entropy production in the thermal bath would not be well-defined).
The operators $\OP{L}_2^\dagger$, $\OP{L}_1^\dagger$, $\OP{L}_0^\dagger$ of the Fokker-Planck equation
\eref{eq:FPEeps} now take the form
\numparts
\begin{eqnarray}
\OP{L}_2^\dagger =
\frac{d\Theta-\tilde{\VEC{v}}^2}{2\Theta^2}\frac{\partial\Theta}{\partial\tilde{t}}\,
\frac{\partial}{\partial\tilde{\F}}
\label{eq:EPSML2}
\, , \\
\OP{L}_1^\dagger =
\tilde{v}_i \frac{\partial}{\partial\tilde{x}_i}
+ \tilde{f}_i \frac{\partial}{\partial\tilde{v}_i}
+ \left(
	\frac{\tilde{f}_i\tilde{v}_i}{\Theta}+ \frac{d\Theta-\tilde{\VEC{v}}^2}{2\Theta^2}\frac{\partial\Theta}{\partial\tilde{x}_i}\tilde{v}_i
  \right)
\frac{\partial}{\partial\tilde{\F}}
\label{eq:EPSML1}
\, , \\
\OP{L}_0^\dagger =
-\tilde{\TENSOR{\gamma}}_{ij} \frac{\partial}{\partial\tilde{v}_i} \tilde{v}_j
- \Theta \, \tilde{\TENSOR{\gamma}}_{ij} \frac{\partial}{\partial\tilde{v}_i}\frac{\partial}{\partial\tilde{v}_j}
\label{eq:EPSML0}
\, .
\end{eqnarray}
\endnumparts
They are completely analogous to \eref{eq:SML0}-\eref{eq:SML2},
except that now the terms related to the functional $\tilde{\F}$
contain polynomials quadratic and cubic in $\tilde{\VEC{v}}$,
which complicate the homogenization procedure considerably.
Details of the corresponding calculation are given in \ref{app:homogenizationEP},
the resulting Fokker-Planck equation reads
\begin{eqnarray}
-\frac{\partial\bar{\rho}}{\partial\tilde{t}} =
	\frac{\partial}{\partial\tilde{x}_i} (\tilde{\TENSOR{\gamma}}^{-1})_{ij}\tilde{f}_j \rho
	- \frac{\partial}{\partial\tilde{x}_i} (\tilde{\TENSOR{\gamma}}^{-1})_{ij} \frac{\partial}{\partial\tilde{x}_j}\Theta\,\rho
\nonumber \\
\qquad\qquad \mbox{}
+ \frac{1}{\Theta}\left( \tilde{f}_i-\frac{\partial\Theta}{\partial\tilde{x}_i} \right)(\tilde{\TENSOR{\gamma}}^{-1})_{ij}\tilde{f}_j\, \frac{\partial\bar{\rho}}{\partial\tilde{\F}}
\nonumber \\
\qquad\qquad \mbox{}
- \frac{\partial}{\partial\tilde{x}_i} (\tilde{\TENSOR{\gamma}}^{-1})_{ij}\left( \tilde{f}_j-\frac{\partial\Theta}{\partial\tilde{x}_j} \right) \frac{\partial\rho}{\partial\tilde{\F}}
\nonumber \\
\qquad\qquad \mbox{}
- \frac{1}{\Theta}\left( \tilde{f}_i-\frac{\partial\Theta}{\partial\tilde{x}_i} \right)(\tilde{\TENSOR{\gamma}}^{-1})_{ij}
	\frac{\partial}{\partial\tilde{x}_j} \Theta\, \frac{\partial\rho}{\partial\tilde{\F}}
\nonumber \\
\qquad\qquad \mbox{}
- \frac{1}{\Theta}\left( \tilde{f}_i-\frac{\partial\Theta}{\partial\tilde{x}_i} \right)(\tilde{\TENSOR{\gamma}}^{-1})_{ij}
	\frac{1}{\Theta}\left( \tilde{f}_j-\frac{\partial\Theta}{\partial\tilde{x}_j} \right) \Theta\, \frac{\partial^2\rho}{\partial\tilde{\F}^2}
\nonumber \\
\qquad\qquad \mbox{}
+ \frac{1}{2\Theta} \frac{\partial\Theta}{\partial\tilde{x}_i}
  \left[ \frac{2}{3}\tilde{\TENSOR{\gamma}}^{-1} + \sum_{l}(\tilde{\TENSOR{\gamma}}+2\tilde{\gamma}^{(l)}\TENSOR{1})^{-1} \right]_{ij}
  \frac{\partial\Theta}{\partial\tilde{x}_j}\,
  \frac{\partial\rho}{\partial\tilde{\F}}
\nonumber \\
\qquad\qquad \mbox{}
- \frac{1}{2\Theta} \frac{\partial\Theta}{\partial\tilde{x}_i}
	\left[ \frac{2}{3}\tilde{\TENSOR{\gamma}}^{-1} + \sum_{l}(\tilde{\TENSOR{\gamma}}+2\tilde{\gamma}^{(l)}\TENSOR{1})^{-1} \right]_{ij}
	\frac{\partial\Theta}{\partial\tilde{x}_j}\,
	\frac{\partial^2 \rho}{\partial\tilde{\F}^2}
\label{eq:homogenizedFPE4EP}
\, ,
\end{eqnarray}
where the $\tilde{\gamma}^{(i)}$ are the eigenvalues of $\tilde{\TENSOR{\gamma}}$.
To write the equivalent Langevin equation we switch back to dimensionful quantities,
\numparts
\begin{eqnarray}
\d\VEC{x}_t =
	\TENSOR{\gamma}^{-1}(\VEC{x}_t) \VEC{f}(\VEC{x}_t,t) \,\d t
	+ \kB T(\VEC{x}_t,t) [\nabla_{\VEC{x}_t}\TENSOR{\gamma}^{-1}(\VEC{x}_t)]\,\d t
\nonumber \\ \qquad\qquad\qquad \mbox{}
	+ \sqrt{2\kB T(\VEC{x}_t,t)}\, \TENSOR{\gamma}^{-1/2}(\VEC{x}_t) \,\d\VEC{W}_t
\label{eq:limitx}
\, , \\[2ex]
\d S^{\mathrm{env}}_t = \frac{1}{T(\VEC{x}_t,t)}\left[ \VEC{f}(\VEC{x}_t,t)- \nabla_{\VEC{x}_t} \kB T(\VEC{x}_t,t) \right]\circ\d\VEC{x}_t
\nonumber \\ \qquad\qquad\qquad \mbox{}
+ \frac{1}{2T} \, (\nabla_{\VEC{x}_t}T)
  \left[ \frac{2}{3}\TENSOR{\gamma}^{-1} + \sum_{l}(\TENSOR{\gamma}+2\gamma^{(l)}\TENSOR{1})^{-1} \right]
  (\nabla_{\VEC{x}_t}T) \,\d t
  \nonumber \\ \qquad\qquad\qquad \mbox{}
+ \frac{1}{\sqrt{T}} \, (\nabla_{\VEC{x}_t}T)
  \left[ \frac{2}{3}\TENSOR{\gamma}^{-1} + \sum_{l}(\TENSOR{\gamma}+2\gamma^{(l)}\TENSOR{1})^{-1} \right]^{1/2}
  \d \hat{\VEC{W}}_t
\label{eq:limitS}
\, .
\end{eqnarray}
\endnumparts
We recover the same dynamics in $\VEC{x}_t$ as before, see \eref{eq:overampedLEDB},
because we considered the exactly identical dynamical process.
The dynamics of entropy production, however, looks quite differently from the one for heat
\eref{eq:overampedHeatDB}. In particular, entropy production consists of two parts,
a regular one (first line of $\d S^{\mathrm{env}}_t$), and a so-called ``anomalous'' contribution \cite{celani2012anomalous}
(second and third line of $\d S^{\mathrm{env}}_t$; note that the Wiener process $\hat{\VEC{W}}_t$
in \eref{eq:limitS} is independent of the Wiener process $\VEC{W}_t$
appearing in the equation \eref{eq:limitx} for $\VEC{x}_t$).
The regular part is expressed as a differential in $\d\VEC{x}_t$, i.e.\ it is determined
by the statistical properties of the process $\VEC{x}_t$.
We point out that it is not simply heat divided by temperature $\VEC{f}\circ\d\VEC{x}_t/T$,
but that for inhomogeneous temperature it contains an additional term.
This expression is consistent with the definition of entropy based on time reversal that one would obtain for the dynamics described by \eref{eq:overampedLEDB}.
For a space-dependent temperature there is also an additional ``anomalous'' entropy.
Its essential property is that it cannot be expressed exclusively as a functional over the trajectory
 $\Traj{\VEC{x}}$, it rather has ``its own'' random fluctuations,
which are completely independent of the fluctuations in $\VEC{x}_t$
(third line of $\d S_t$).
As a consequence of the coarse-graining associated with the asymptotic limit $\tau_v/\tau_x \to 0$,
the trajectory $\Traj{\VEC{x}}$ alone does not contain sufficient
information to fully specify entropy production in an inhomogeneous environment.
The average ``anomalous'' entropy production corresponds to the drift term in the second line
of $\d S^{\mathrm{env}}_t$, and is non-negative.
Note that for $\TENSOR{\gamma}=\gamma_0\TENSOR{1}$, where $\TENSOR{1}$ is the identity tensor,
with a scalar friction
coefficient $\gamma_0$, this term reads
$\frac{1}{T}\frac{d+2}{6\gamma_0}(\nabla_{\VEC{x}_t}T)^2$.
Its physical origin and properties are discussed in detail in
\cite{celani2012anomalous,bo2013entropic,bo2014entropy,marino2016entropy,miyazaki2018entropy}.
We emphasize that, if temperature is constant in space, the limiting procedure is regular and one simply finds that entropy is given by the usual (minus) heat divided by temperature recovering \eref{eq:DeltaS} with the expected Stratonovich product.

\section{Discussion and Conclusions}
The generalized Langevin equation \eref{eq:GLE} and the associated functionals
along its trajectories do not suffer from any ambiguity related to interpretation of stochastic integrals, because
the generated position and velocity processes are sufficiently regular.
Exploiting this uniqueness of integrals, we here applied a multiscale procedure to the GLE
and its functionals to find out which noise interpretations
of the overdamped Langevin equation and its thermodynamic functionals are consistent with the systematic
white-noise and small-mass limits.
The most important finding is that heat- and work-like functionals need to be interpreted
in Stratonovich sense, no matter which kind of multiplicative noise is present in the
underlying overdamped Langevin equation. This is in agreement with physical arguments
based on consistency of the energy balance \cite{sekimoto2010stochastic}.
We remark that, before performing the limiting procedure on the heat functional
\eref{eq:QGLE} we have split off the boundary term corresponding to the changes
in kinetic energy of the particle. Its average contribution is proportional to the
change in temperature $d \kB\Delta T/2$ (where $d$ is the dimensionality of the system).
As pointed out in \cite{bo2013entropic} this contribution plays a relevant role when
temperature is changing in time and must be kept into account to properly define adiabatic transformations, which have then been experimentally realized \cite{martinez2015adiabatic,martinez2016brownian}.
The average contribution of the boundary term has been computed for two cases studies
also in \cite{arold2018heat} and its probability distribution in \cite{garcia2018heat}

In addition to heat and work, we also discussed the case of entropy production
\cite{celani2012anomalous}, which, in presence
of spatial temperature variations, cannot
be expressed as a functional over the overdamped stochastic trajectory only.
It rather features  an
additional ``anomalous'' contribution,
which is related to dissipation
due to heat transport by the (hidden) velocity degrees of freedom exchanged with different thermostats
\cite{celani2012anomalous,spinney2012entropy,bo2014entropy,marino2016entropy,bo2017multiple,ge2018anomalous,birrell2018entropy}.

We here presented these calculations having specifically in mind a Brownian particle
suspended in an aqueous solution. Generalizations to several (interacting) Brownian particles
are straightforward. In such systems, the intrinsic time scale of the thermal bath
is the fastest time scale in the system, followed by the relaxation time scale of the velocity degrees of freedom
of the Brownian particle. Accordingly, we performed the white-noise limit first,
and then the small-mass limit.
However, the GLE, which in its most basic form was
first introduced by Mori in \cite{mori1965transport}, has subsequently been used to model many systems
in statistical and biological physics \cite{goychuk2012viscoelastic,lysy2016model},
including both normal and anomalous diffusion, and the motion of active matter \cite{sevilla2018non}.
In general, the separation of time scales might be different in other physical systems, distinct from a
(passive) Brownian particle, such that other orderings of the
above limits might become relevant. While it is known that the ordering of limits 
affects the limiting dynamics \cite{kupferman2004ito}, we are not aware of any results for functionals.

Further very interesting generalizations of our analysis based on the model
\eref{eq:GLE}-\eref{eq:etadot} include the case of many particles
subject to a matrix-valued
friction kernel $\TENSOR{\kappa}$ with different relaxation times, and likewise for
the correlations of the colored noise $\VEC{\eta}$.
This would allow for analyzing, e.g., the situation
in which different particles are in contact with heat baths of different physical character.
The heat baths  can even have different temperatures
by appropriate generalization of $T$ or choice of $\TENSOR{\sigma}$.
Other non-equilibrium systems of great interest, in particular in the context of active matter,
are Brownian particles driven by colored noise processes $\VEC{\eta}$ which are themselves
out of equilibrium. We expect non-trivial contributions to the dynamics \cite{bo2013white}
and thermodynamics (i.e.\ functionals),
resulting in implicit discretization rules which are more complicated than just the
common rules such as the It\^o, Stratonovich and anti-It\^o rule \cite{pavliotis2008multiscale}.

All these ideas are left for future explorations.

\ack
The authors gratefully acknowledge stimulating discussions with Peter H\"{a}nggi.
\changed{SB and RE thank Antonio Celani and Erik Aurell for inspiring collaborations on
a variety of problems involving multi-scale techniques.}
SB and RE acknowledge financial support from the Swedish Research Council
(Vetenskapsr{\aa}det) under the grants No.~638-2013-9243 and No.~2016-05412. 

\appendix
\section{Homogenization of the Fokker-Planck equation}
\label{app:HFPE}
In this Appendix we give the most relevant technical details for the homogenization procedure
used to calculate the white-noise and small-mass limits.
We start from a general form of the Langevin equation and its equivalent Fokker-Planck
equation, which includes both these cases from the main text,
see Sections~\ref{sec:WNlimit} and \ref{sec:SMlimit}.
Collecting all slow variables in the vector $\VEC{X}$, all fast variables in $\VEC{Y}$,
and denoting the functional as $\F$ as before, the general form of the Langevin equation reads
\numparts
\begin{eqnarray}
\d\VEC{X}_t = \VEC{U}(\VEC{X}_t,t) \,\d t + \frac{1}{\epsilon}\TENSOR{S}(\VEC{X}_t,t)\VEC{Y}_t \,\d t
\label{eq:gdX}
\, , \\
\d\VEC{Y}_t = \frac{1}{\epsilon}\VEC{V}(\VEC{X}_t,t) \,\d t
	- \frac{1}{\epsilon^2}\TENSOR{G}(\VEC{X}_t,t)\VEC{Y}_t \,\d t
	+ \frac{1}{\epsilon}\sqrt{2}\,\TENSOR{D}^{1/2}(\VEC{X}_t,t)\cdot\d\VEC{W}_t
\label{eq:gdY}
\, , \\
\d\F_t = R_0(\VEC{X}_t,\VEC{Y}_t,t) \,\d t + \frac{1}{\epsilon}R_1(\VEC{X}_t,\VEC{Y}_t,t) \,\d t
\label{eq:gdF}
\, .
\end{eqnarray}
\endnumparts
Here, $R_0$, $R_1$ are scalar functions, $\VEC{U}$, $\VEC{V}$ are vector-valued functions and
$\TENSOR{S}$, $\TENSOR{G}$, and $\TENSOR{D}$ are matrix-valued functions of the indicated arguments;
their dimensions are fixed by the number of slow and fast variables, respectively.
The small parameter $\epsilon$ represents the
(square root of the) ratio between the fastest and slowest time scale
in the system.
We assume that the explicit time dependencies in all the functions
$R_0$, $R_1$, $\VEC{U}$, $\VEC{V}$, $\TENSOR{S}$, $\TENSOR{G}$, and $\TENSOR{D}$
occur on the slowest time scale only.

We can write the Fokker-Planck equation for the density
$\rho = \rho(t,\VEC{X},\VEC{Y},\F)$, which is equivalent to the Langevin equations
\eref{eq:gdX}-\eref{eq:gdF}, in the form
\begin{equation}
\label{eq:gFPE}
\frac{\partial\rho}{\partial t} =
-\left(
	\OP{L}_2^\dagger + \frac{1}{\epsilon}\OP{L}_1^\dagger + \frac{1}{\epsilon^2}\OP{L}_0^\dagger
\right)\rho
\end{equation}
with the operators
\numparts
\begin{eqnarray}
\OP{L}_2^\dagger = \frac{\partial}{\partial X_i} U_i + R_0 \frac{\partial}{\partial\F}
\label{eq:gL0}
\, , \\
\OP{L}_1^\dagger = \frac{\partial}{\partial X_i}S_{ij}Y_j + V_i \frac{\partial}{\partial Y_i} + R_1 \frac{\partial}{\partial\F}
\label{eq:gL1}
\, , \\
\OP{L}_0^\dagger = - G_{ij}\frac{\partial}{\partial Y_i}Y_j - D_{ij}\frac{\partial}{\partial Y_i}\frac{\partial}{\partial Y_j}
\label{eq:gL2}
\, ,
\end{eqnarray}
\endnumparts
where we skipped the arguments of the functions and switched to index notation
with summation over repeated indices being understood.
The ``translations'' of this general Fokker-Planck equation into the specific
forms \eref{eq:WNL0}-\eref{eq:WNL2} and \eref{eq:SML0}-\eref{eq:SML2}
used for the white-noise and the small-mass limit, respectively, are provided in
table \ref{tab:map}.
\begin{table}
\caption{\label{tab:map}
``Translation key'' to map the general form \eref{eq:gL0}, \eref{eq:gL1}, \eref{eq:gL2}
to the specific Fokker-Planck equations \eref{eq:WNL0}, \eref{eq:WNL1}, \eref{eq:WNL2}
and \eref{eq:SML0}, \eref{eq:SML1}, \eref{eq:SML2} from the  main text.}
\begin{indented}
\item[]\begin{tabular}{@{}lll}
\br
& \textbf{white-noise limit} & \textbf{small-mass limit}
\\
& eqs.~\eref{eq:WNL0}, \eref{eq:WNL1}, \eref{eq:WNL2}
& eqs.~\eref{eq:SML0}, \eref{eq:SML1}, \eref{eq:SML2}
\\
\mr
$\VEC{X}$ & $(\tilde{\VEC{x}},\tilde{\VEC{v}})$ & $\tilde{\VEC{x}}$
\\[2ex]
$\VEC{Y}$ & $(\tilde{\VEC{y}},\tilde{\VEC{\eta}})$ & $\tilde{\VEC{v}}$
\\[2ex]
$\VEC{U}(\VEC{X},t)$
&
$\left(
	\sqrt{\frac{\tau_x}{\tau_v}}\tilde{\VEC{v}},
	\sqrt{\frac{\tau_x}{\tau_v}}\tilde{\VEC{f}}
\right)$
&
$\VEC{0}$
\\[2ex]
$\VEC{V}(\VEC{X},t)$
&
$\left( \sqrt{\frac{\tau_x}{\tau_v}}\TENSOR{h}\tilde{\VEC{v}},\VEC{0} \right)$
&
$\tilde{\VEC{f}}$
\\[1ex]
$\TENSOR{S}(\VEC{X},t)$
&
$\left( \begin{array}{cc}
\TENSOR{0} & \TENSOR{0} \\
-\sqrt{\frac{\tau_x}{\tau_v}}\TENSOR{g} & \sqrt{\frac{\tau_x}{\tau_v}}\sqrt{2\Theta}\,\TENSOR{\sigma}
\end{array}\right)$
&
$\TENSOR{1}$
\\[2ex]
$\TENSOR{G}(\VEC{X},t)$
&
$\left( \begin{array}{cc}
\TENSOR{1} & \TENSOR{0} \\
\TENSOR{0} & \TENSOR{1}
\end{array}\right)$
&
$\TENSOR{g}\TENSOR{h}$
\\[2ex]
$\TENSOR{D}(\VEC{X},t)$
&
$\frac{1}{2}\left( \begin{array}{cc}
\TENSOR{0} & \TENSOR{0} \\
\TENSOR{0} & \TENSOR{1}
\end{array}\right)$
&
$\Theta\,\TENSOR{\sigma}\TENSOR{\sigma}\TRANS$
\\[2ex]
$R_0(\VEC{X},\VEC{Y},t)$
&
$\tilde{r} + \sqrt{\frac{\tau_x}{\tau_v}}\, \tilde{\VEC{q}}\cdot\tilde{\VEC{v}}$
&
$\tilde{r}$
\\[2ex]
$R_1(\VEC{X},\VEC{Y},t)$
&
$0$
&
$\tilde{\VEC{q}}\cdot\tilde{\VEC{v}}$
\\[0ex]
\br
\end{tabular}
\end{indented}
\end{table}
\subsection{The homogenization procedure}
The homogenization procedure \cite{pavliotis2008multiscale,bo2017multiple}
consists in a systematic perturbation expansion in the small parameter $\epsilon$, which
characterizes the separation of time scales present in the system
\footnote{\changed{
For a pedagogical introduction to the multiple time-scale approach
in the context of Fokker-Planck equations
we recommend Ref.~\cite{bocquet1997high}. Bocquet uses an expansion procedure equivalent to ours,
with some (minor) technical differences on how the solutions of the higher-order equations
(i.e.~Eqs.~\eref{eq:order-1} and \eref{eq:order0} in our case)
are treated.}}.
Accordingly, we introduce time variables $\theta$ and $\tau$ on the fastest and slowest scales, respectively,
and an intermediate time scale $\vartheta$,
\begin{equation}
\label{eq:tscales}
\theta = \epsilon^{-2}t
\, , \quad
\vartheta = \epsilon^{-1}t
\, , \quad
\tau = t
\, .
\end{equation}
Then, we expand the density $\rho$ in powers of $\epsilon$,
\begin{equation}
\label{eq:rhoexp}
\rho = \rho_0 + \epsilon \rho_1 + \epsilon^2 \rho_2 + \ldots
\, ,
\end{equation}
where all $\rho_i=\rho_i(\vartheta,\tau,\VEC{X},\VEC{Y},\F)$ are a priori assumed to be
functions of all variables and the two slower time scales independently.
Since we are not interested in the relaxation processes on the fastest time scale $\theta$,
but rather in the situation in which these fastest scales reached their ``equilibrated''
stationary state, we assume the $\rho_i$ to be independent of $\theta$.
Plugging the expansion \eref{eq:rhoexp} into the Fokker-Planck equation \eref{eq:gFPE},
and using that the time-derivative turns into
$\partial/\partial t = \epsilon^{-2}\partial/\partial\theta + \epsilon^{-1}\partial/\partial\vartheta + \partial/\partial\tau$
according to \eref{eq:tscales} (and the assumption of the three time scales to be independent),
we collect together all terms with the same power in $\epsilon$
to find the hierarchy of equations
\numparts
\begin{eqnarray}
\OP{L}_0^\dagger\rho_0 = 0
\label{eq:order-2}
\, , \\
\OP{L}_0^\dagger\rho_1 = -\frac{\partial\rho_0}{\partial\vartheta} - \OP{L}_1^\dagger\rho_0
\label{eq:order-1}
\, , \\
\OP{L}_0^\dagger\rho_2 = -\frac{\partial\rho_0}{\partial\tau}-\frac{\partial\rho_1}{\partial\vartheta} - \OP{L}_2^\dagger\rho_0 - \OP{L}_1^\dagger\rho_1
\label{eq:order0}
\, .
\end{eqnarray}
\endnumparts

From \eref{eq:gL2} we see that the solution to \eref{eq:order-2} can be written as
\begin{equation}
\label{eq:rho0}
\rho_0 = w(\VEC{Y}|\VEC{X},\tau) \bar{\rho}_0(\vartheta,\tau,\VEC{X},\F)
\, ,
\end{equation}
where the density $w(\VEC{Y}|\VEC{X},\tau)$ fulfills
\begin{equation}
\OP{L}_0^\dagger w = 0
\, , \quad
\int \d\VEC{Y} w = 1
\, ,
\end{equation}
i.e.\ it is the stationary density of the fast variables $\VEC{Y}$
conditioned on the slow variable $\VEC{X}$ and slow time $\tau$.
Due to the simple form of \eref{eq:gL2}, we can calculate $w$ easily and obtain the Gaussian
\begin{equation}
\label{eq:gw}
w(\VEC{Y}|\VEC{X},\tau) = \frac{1}{(2\pi)^{n/2}\sqrt{\det\TENSOR{\Sigma}}}\,e^{-\frac{1}{2}Y_i(\TENSOR{\Sigma}^{-1})_{ij}Y_j}
\, ,
\end{equation}
where $n$ is the dimension of the vector $\VEC{Y}$.
The symmetric covariance matrix $\Sigma$ is the solution of the Lyapunov equation
\begin{equation}
\label{eq:glyapunov}
\TENSOR{G}\TENSOR{\Sigma} + \TENSOR{\Sigma}\TENSOR{G}\TRANS = 2\TENSOR{D}
\, .
\end{equation}
Note that in general it depends on $\VEC{X}$ and $\tau$ via the functional dependencies of
$\TENSOR{G}$, $\TENSOR{D}$,
such that also the covariances
\begin{equation}
\overline{Y_i Y_j} = \int \d\VEC{Y}\, Y_i Y_j w = \Sigma_{ij}
\end{equation}
depend explicitly on the slow scales $\VEC{X}$ and $\tau$
(the overline denotes the average over the fast variables $\VEC{Y}$ weighed with their stationary density $w$),
justifying our notation of a density $w(\VEC{Y}|\VEC{X},\tau)$ conditioned on $\VEC{X}$ and $\tau$.

We proceed with the next equation \eref{eq:order-1} by employing the so-called
solvability condition \cite{pavliotis2008multiscale}, which states that the right-hand side inhomogeneity
has to be orthogonal to the nullspace of the operator $\OP{L}_0$ adjoint to $\OP{L}_0^\dagger$,
otherwise there is no non-trivial solution.
From \eref{eq:gL2} we see that the constant functions are contained in the nullspace of $\OP{L}_0$,
such that the solvability condition reads
$\int\d\VEC{Y}\left( \frac{\partial\rho_0}{\partial\vartheta} + \OP{L}_1^\dagger\rho_0\right)=0$.
Using the explicit form \eref{eq:gL1} and the solution \eref{eq:rho0}, it is straightforward to show
that the solvability condition reduces to
$\frac{\partial\bar{\rho}_0}{\partial\vartheta}=-\overline{R_1}\,\frac{\partial\bar{\rho}_0}{\partial\F}$.
Since the equation \eref{eq:order-1} represents contribution of order $\epsilon^{-1}$ we require them
to vanish to guarantee a well-defined $\epsilon\to 0$ limit (so-called centering condition).
In our case, we therefore have to restrict the function $R_1$ to be \emph{odd} in the fast variables $\VEC{Y}$.
Then, we have $\overline{R_1}=0$, and we find that also $\frac{\partial\bar{\rho}_0}{\partial\vartheta}=0$, i.e.\ that
the reduced density $\bar{\rho}_0$ is independent on the intermediate time scale $\vartheta$.

Applying the solvability condition in an analogous way to the third equation \eref{eq:order0} of
our equation hierarchy, we find
\begin{eqnarray}
-\left( \frac{\partial\bar{\rho}_0}{\partial\tau} + \frac{\partial\bar{\rho}_1}{\partial\vartheta} \right)
& = &
\int \d\VEC{Y} \left( \OP{L}_2^\dagger\rho_0 + \OP{L}_1^\dagger\rho_1 \right)
\nonumber \\
& = &
\frac{\partial}{\partial X_i}U_i\bar{\rho}_0
+ \overline{R_0}\, \frac{\partial\bar{\rho}_0}{\partial\F}
+ \frac{\partial}{\partial X_i} S_{ij} J_{Y_j}
+ \frac{\partial}{\partial\F} J_{R_1}
\label{eq:SCorder0}
\, .
\end{eqnarray}
In the second equality, we have used the solution \eref{eq:rho0} for $\rho_0$,
the explicit expressions for the operators
$\OP{L}_2^\dagger$ and $\OP{L}_1^\dagger$ from \eref{eq:gL0} and \eref{eq:gL1}, and we have
introduced the definition of the current-like quantity
\begin{equation}
\label{eq:Jpi}
J_{\pi(\VEC{Y})} = \int \d\VEC{Y}\,\pi(\VEC{Y})\rho_1
\end{equation}
for (polynomial) functions $\pi(\VEC{Y})$ of the fast variables.
It turns out that the result \eref{eq:SCorder0} is
sufficient to derive a Fokker-Planck equation for the reduced density
$\bar{\rho} = \int\d\VEC{Y}\,\rho$ in slow variables only
(being the reason that we do not need to go to higher orders in the equation hierarchy \eref{eq:order-2}-\eref{eq:order0}). 
Our goal is to identify the lowest order
contributions to the dynamics of $\bar{\rho}$ which
survive in the limit $\epsilon \to 0$.
Indeed, from the definition \eref{eq:tscales} of the three time scales and the
expansion \eref{eq:rhoexp} we find
$\frac{\partial\bar{\rho}}{\partial t}=\frac{\partial\bar{\rho}_0}{\partial\tau}
+ \frac{\partial\bar{\rho}_1}{\partial\vartheta} + \cal{O}(\epsilon)$
and $\bar{\rho}=\bar{\rho}_0 + \cal{O}(\epsilon)$.
Hence, in the asymptotic limit $\epsilon \to 0$ we obtain from \eref{eq:SCorder0}
\begin{equation}
\label{eq:homogenizedFPEC}
-\frac{\partial\bar{\rho}}{\partial t} =
\frac{\partial}{\partial X_i}U_i\bar{\rho}
+ \frac{\partial}{\partial X_i} S_{ij} J_{Y_j}
+ \overline{R_0}\, \frac{\partial\bar{\rho}}{\partial\F}
+ \frac{\partial}{\partial\F} J_{R_1}
\, .
\end{equation}
As a Gaussian integral over $w$, the average $\overline{R_0}$ is straightforward to calculate
if $R_0$ contains polynomials in $\VEC{Y}$, as it is typically the case.
The remaining task consists in expressing
$J_{Y_i}$ and $J_{R_1}$, which contain the a priori unknown solution $\rho_1$
of the equation \eref{eq:order-1},
in terms of $\bar{\rho}=\bar{\rho}_0 + \cal{O}(\epsilon)$
in lowest order $\epsilon$.

\subsection{Heat- and work-like functionals linear in $\VEC{Y}$}
For explicit results we first focus on the case, which is relevant for the heat and work
functionals, i.e.
\numparts
\begin{eqnarray}
R_0 = R_0(\VEC{X},t)
\label{eq:gR0}
\, , \\
R_1 = Q_i(\VEC{X},t)Y_i
\label{eq:gR1}
\, ,
\end{eqnarray}
\endnumparts
with functions $Q_i$ depending on slow scales only.
As demonstrated in \ref{app:JYi}, we find
\numparts
\begin{eqnarray}
J_{Y_i} = - (\TENSOR{G}^{-1})_{ij} \left[
\frac{\partial}{\partial X_l}\Sigma_{jk}(\TENSOR{S}\TRANS)_{kl} - V_j + \Sigma_{jk}Q_k \frac{\partial}{\partial\F}
\right] \bar{\rho}
\label{eq:JYiHW}
\, , \\
J_{R_1} = Q_i J_{Y_i}
\, .
\end{eqnarray}
\endnumparts
Hence, the final Fokker-Planck equation reads
\begin{eqnarray}
-\frac{\partial\bar{\rho}}{\partial t} =
\frac{\partial}{\partial X_i} \left[
U_i + S_{ij}(\TENSOR{G}^{-1})_{jk}V_k
\right] \bar{\rho}
- \frac{\partial}{\partial X_i} S_{ij}(\TENSOR{G}^{-1})_{jk} \frac{\partial}{\partial X_m} \Sigma_{kl}(\TENSOR{S}\TRANS)_{lm}\,\bar{\rho}
\nonumber \\
\qquad\qquad \mbox{}
+ \left[ R_0 + Q_i(\TENSOR{G}^{-1})_{ij}V_j \right] \frac{\partial\bar{\rho}}{\partial\F}
\nonumber \\
\qquad\qquad \mbox{}
- \frac{\partial}{\partial X_i} S_{ij}(\TENSOR{G}^{-1})_{jk}\Sigma_{kl}Q_l \frac{\partial\bar{\rho}}{\partial\F}
- Q_i(\TENSOR{G}^{-1})_{ij} \frac{\partial}{\partial X_l} \Sigma_{jk}(\TENSOR{S}\TRANS)_{kl} \frac{\partial\bar{\rho}}{\partial\F}
\nonumber \\
\qquad\qquad \mbox{}
- Q_i(\TENSOR{G}^{-1})_{ij}\Sigma_{jk}Q_k \frac{\partial^2\bar{\rho}}{\partial\F^2}
\, .
\end{eqnarray}
From this general result we can easily read off the two homogenized Fokker-Planck equations
\eref{eq:underdampedFPE} and \eref{eq:overdampedFPE} for the white-noise and small-mass limit,
respectively, by applying the mapping provided in Table \ref{tab:map}.

\subsection{Entropy production (quadratic and cubic in $\VEC{Y}$)}
\label{app:homogenizationEP}
Our second explicit example is the entropy production from Section \ref{sec:EP}.
We focus on the physically relevant case when the fluctuation-dissipation relation holds,
i.e.\ in the general notation of this Appendix,
\begin{equation}
\label{eq:DEP}
	\TENSOR{D}=\Theta\TENSOR{G}=\Theta\TENSOR{G}\TRANS
\, ,
\end{equation}
where we split off the (in general space- and time-dependent) temperature $\Theta$.
As a consequence, $\TENSOR{\Sigma} = \Theta\TENSOR{1}$ (see \eref{eq:glyapunov}),
such that the stationary density \eref{eq:gw} for the fast variables simplifies to
\numparts
\begin{equation}
\label{eq:wEP}
w = \frac{1}{(2\pi\Theta)^{n/2}}\,e^{-\frac{\VEC{Y}^2}{2\Theta}}
\, ,
\end{equation}
with its variance at $\VEC{X}$ and $t$ being set by the local temperature,
\begin{equation}
\label{eq:YvarTheta}
\overline{Y_i Y_j}=\Theta\delta_{ij} = \Theta(\VEC{X},t)\,\delta_{ij}
\, .
\end{equation}
\endnumparts
Moreover, we restrict ourselves to the specific structure of linear, quadratic and cubic terms
in the fast variables as they appear
in the functional for the entropy production \eref{eq:dFEP},
keeping, in particular, the differences between $\VEC{Y}^2$ and
their variance,
\numparts
\begin{eqnarray}
R_0 = \left( n\Theta - \VEC{Y}^2 \right) R(\VEC{X},t)
\, , \\
R_1 = Q_i(\VEC{X},t) Y_i + \left( n\Theta - \VEC{Y}^2 \right) Y_i \, P_i(\VEC{X},t)
\label{eq:R1EP}
\, .
\end{eqnarray}
\endnumparts
In case of the entropy production, the functions $R(\VEC{X},t)$,
$\VEC{Q}(\VEC{X},t)$ and $\VEC{P}(\VEC{X},t)$ read
\begin{equation}
\label{eq:RQPEP}
R = \frac{1}{2\Theta^2}\frac{\partial\Theta}{\partial{t}}
\, , \quad
\VEC{Q} = \frac{\tilde{\VEC{f}}}{\Theta}
\, , \quad
\VEC{P} = \frac{1}{2\Theta^2}\frac{\partial\Theta}{\partial\VEC{X}}
\, .
\end{equation}
Note that all indicated explicit time dependencies are assumed to
occur on the slowest scale $\tau$.

We have to use these expressions for $R_0$ and $R_1$ to evaluate
the average $\overline{R_0}$ and the integral $J_{R_1}$ appearing
in \eref{eq:homogenizedFPEC}. We first notice that (see \eref{eq:YvarTheta})
\numparts
\begin{equation}
\overline{R_0} = n\Theta - \overline{Y_i Y_i} = n\Theta - n\Theta = 0
\, ,
\end{equation}
and that we can write
\begin{equation}
J_{R_1} = Q_i J_{Y_i} + P_i \left( n\Theta J_{Y_i} - J_{Y_j Y_j Y_i} \right)
\, .
\end{equation}
\endnumparts
The integral $J_{Y_i}$ is calculated in \ref{app:JYi} (see \eref{eq:JYiEPres}),
\begin{equation}
\label{eq:JYiEP}
J_{Y_i} = (\TENSOR{G}^{-1})_{ij} \left[
V_j - \frac{\partial}{\partial X_k}\Theta(\TENSOR{S}\TRANS)_{jk} - \Theta \left( Q_j - 2\Theta P_j \right) \frac{\partial}{\partial\F}
\right] \bar{\rho}
\, .
\end{equation}
The calculation of $J_{Y_j Y_j Y_i}$ is sketched in \ref{app:JYjYjYi}.
We find (cf.~\eref{eq:JYjYjYires})
\begin{equation}
J_{Y_j Y_j Y_i} =
\Theta (n+2)J_{Y_i}
- M_{ij} \left[ \Theta \frac{\partial\Theta}{\partial X_k} S_{kj} - 2\Theta^3 P_j \frac{\partial}{\partial\F} \right] \bar{\rho}
\, ,
\end{equation}
where $\TENSOR{M}$ is defined in \eref{eq:M}. Hence,
\begin{eqnarray}
J_{R_1} =
(Q_i - 2\Theta P_i) J_{Y_i}
+P_i M_{ij} \left[ \Theta \frac{\partial\Theta}{\partial X_k} S_{kj} - 2\Theta^3 P_j \frac{\partial}{\partial\F} \right] \bar{\rho}
\, .
\end{eqnarray}
Plugging all these results into \eref{eq:homogenizedFPEC} we finally obtain
the homogenized Fokker-Planck equation
\begin{eqnarray}
-\frac{\partial\bar{\rho}}{\partial t} =
	\frac{\partial}{\partial X_i} \left[ U_i + S_{ij}(\TENSOR{G}^{-1})_{jk}V_k \right] \bar{\rho}
	- \frac{\partial}{\partial X_i} S_{ij}(\TENSOR{G}^{-1})_{jk} \frac{\partial}{\partial X_l} (\TENSOR{S}\TRANS)_{kl}\Theta\,\bar{\rho}
\nonumber \\
\qquad\qquad \mbox{}
+ (Q_i-2\Theta P_i)(\TENSOR{G}^{-1})_{ij}V_j\, \frac{\partial\bar{\rho}}{\partial\F}
\nonumber \\
\qquad\qquad \mbox{}
- \frac{\partial}{\partial X_i} S_{ij}(\TENSOR{G}^{-1})_{jk}(Q_k-2\Theta P_k) \Theta \frac{\partial\bar{\rho}}{\partial\F}
\nonumber \\
\qquad\qquad \mbox{}
- (Q_i-2\Theta P_i)(\TENSOR{G}^{-1})_{ij} \frac{\partial}{\partial X_k} (\TENSOR{S}\TRANS)_{jk}\Theta \frac{\partial\bar{\rho}}{\partial\F}
\nonumber \\
\qquad\qquad \mbox{}
- \Theta (Q_i-2\Theta P_i)(\TENSOR{G}^{-1})_{ij}(Q_j-2\Theta P_j) \frac{\partial^2\bar{\rho}}{\partial\F^2}
\nonumber \\
\qquad\qquad \mbox{}
+ P_i M_{ij} (\TENSOR{S}\TRANS)_{jk} \frac{\partial\Theta}{\partial X_k}\Theta \frac{\partial\bar{\rho}}{\partial\F}
\nonumber \\
\qquad\qquad \mbox{}
- 2\Theta^3 P_i M_{ij} P_j \frac{\partial^2\bar{\rho}}{\partial\F^2}
\, .
\end{eqnarray}
Applying the translation from Table \ref{tab:map} for the small-mass limit (with
$\TENSOR{h}\TENSOR{g}=\TENSOR{\sigma}\TENSOR{\sigma}\TRANS=\TENSOR{\gamma}$),
the specifications \eref{eq:RQPEP} for entropy production and the definition \eref{eq:M}, we obtain the
Fokker-Planck equation \eref{eq:homogenizedFPE4EP} given in the main text.

\section{The integrals $J_{\pi(\VEC{Y})}$}
We here briefly describe how to calculate the integrals
\eref{eq:Jpi},
\begin{equation}
J_{\pi(\VEC{Y})}=\int\d\VEC{Y}\,\pi(\VEC{Y})\rho_1
\, ,
\end{equation}
for a polynomial function $\pi(\VEC{Y})$ in $\VEC{Y}$
without explicitly solving for $\rho_1$ \cite{marino2016entropy}
\footnote{An alternative way consists in (formally) solving
\eref{eq:order-1} for $\rho_1$ and then evaluating the integral \cite{bo2017multiple}.}.
The central idea is to multiply \eref{eq:order-1}
(remembering that $\frac{\partial\rho_0}{\partial\vartheta}=0$)
by a polynomial $\Pi(\VEC{Y})$ in $\VEC{Y}$ (usually different from $\pi(\VEC{Y}$)
and integrating the left-hand side according to
$\int\d\VEC{Y}\,\Pi(\VEC{Y})\OP{L}_0^\dagger\rho_1=\int\d\VEC{Y}\,[\OP{L}_0\Pi(\VEC{Y})]\rho_1=J_{\OP{L}_0\Pi(\VEC{Y})}$
(see \eref{eq:gL2}), while evaluating the corresponding integral
$\int\d\VEC{Y}\,\Pi(\VEC{Y})\OP{L}_1^\dagger\rho_0$ on the right-hand side using the
known solution \eref{eq:rho0} for $\rho_0$.
In this way, we obtain
\begin{equation}
\label{eq:JL0Y}
J_{\OP{L}_0\Pi(\VEC{Y})} = -\int\d\VEC{Y}\,\Pi(\VEC{Y})\OP{L}_1^\dagger\rho_0
\end{equation}
for yet another polynomial $\OP{L}_0\Pi(\VEC{Y})$ in explicit form.
The trick is to find an appropriate ansatz for $\Pi(\VEC{Y})$,
such that $\OP{L}_0\Pi(\VEC{Y})$ equals to or contains
the desired function $\pi(\VEC{Y})$.
In the following two Sections, we will sketch this calculation for
$\pi(\VEC{Y})=Y_i$ and $\pi(\VEC{Y})=Y_j Y_j Y_i$.

\subsection{Calculation of $J_{Y_i}$}
\label{app:JYi}
In order to calculate $J_{Y_i}$ we choose $\Pi(\VEC{Y})=(\TENSOR{G}^{-1})_{ij}Y_j$,
because then $\OP{L}_0\Pi(\VEC{Y})=Y_i$,
as we find easily from \eref{eq:gL2}.
Hence
\begin{eqnarray}
J_{Y_i}
& = & -\int\d\VEC{Y}\,(\TENSOR{G}^{-1})_{ij}Y_j\OP{L}_1^\dagger\rho_0
\nonumber \\
& = & -(\TENSOR{G}^{-1})_{ij}
	\left[
		\frac{\partial}{\partial X_k} S_{kl}\Sigma_{lj} - V_j + \overline{Y_j R_1} \frac{\partial}{\partial\F}
	\right] \bar{\rho}_0
\end{eqnarray}
where the second line follows after direct integration using the explicit form
\eref{eq:gL1} of the operator $\OP{L}_1^\dagger$ and the solution \eref{eq:rho0}
for $\rho_0$.
Using the general form \eref{eq:R1EP} for $R_1$ we obtain
\begin{equation}
\overline{Y_j R_1} =
\Sigma_{jk} Q_k
+ \left( n\Theta \Sigma_{jk} - \Sigma_{ii}\Sigma_{jk} - 2\Sigma_{ij}\Sigma_{ik} \right) P_k
\, .
\end{equation}
For heat- and work-like functionals we have $P_i \equiv 0$, i.e.\
\numparts
\begin{equation}
\label{eq:JYiHWres}
J_{Y_i}
= -(\TENSOR{G}^{-1})_{ij}
	\left[
		\frac{\partial}{\partial X_k} S_{kl}\Sigma_{lj} - V_j + \Sigma_{jk}Q_k \frac{\partial}{\partial\F}
	\right] \bar{\rho}_0
\, ,
\end{equation}
while $\Sigma_{ij}=\Theta\delta_{ij}$ for the entropy production functional, i.e.\
\begin{equation}
\label{eq:JYiEPres}
J_{Y_i}
= -(\TENSOR{G}^{-1})_{ij}
	\left[
		\frac{\partial}{\partial X_k} \Theta S_{kj} - V_j + \left( \Theta Q_j - 2\Theta^2 P_j \right) \frac{\partial}{\partial\F}
	\right] \bar{\rho}_0
\, .
\end{equation}
\endnumparts
The first expression is the same one as given in \eref{eq:JYiHW},
the second expression agrees with \eref{eq:JYiEP}.

\subsection{Calculation of $J_{Y_j Y_j Y_i}$}
\label{app:JYjYjYi}
We need the integral $J_{Y_j Y_j Y_i}$ only for the entropy production.
In the following, we thus
focus on the specification given by eqs.~\eref{eq:DEP}-\eref{eq:YvarTheta}.
It turns out that the appropriate ansatz for $\Pi(\VEC{Y})$ is
$\Pi(\VEC{Y})=A_{ijkl}Y_j Y_k Y_l$ with a fourth-order tensor $A_{ijkl}$.
Then, we obtain from \eref{eq:gL2} (with \eref{eq:DEP})
\begin{eqnarray}
\OP{L}_0 A_{ijkl}Y_j Y_k Y_l =
A_{ijkl} \left(
	G_{jm} Y_m Y_k Y_l + G_{km} Y_m Y_j Y_l + G_{lm} Y_m Y_j Y_k
\right)
\nonumber \\ \qquad\qquad\qquad\qquad  \mbox{}
- 2\Theta A_{ijkl} \left(
	G_{jk} Y_l + G_{jl} Y_k + G_{kl} Y_j
\right)
\, ,
\end{eqnarray}
and construct $A_{ijkl}$ in a way that the combination of third-order polynomials in $\VEC{Y}$
reduces to $Y_j Y_j Y_i$, i.e.\
$A_{ijkl} \left( G_{jm} Y_m Y_k Y_l + G_{km} Y_m Y_j Y_l + G_{lm} Y_m Y_j Y_k \right) = Y_j Y_j Y_i$.
This is achieved by
\begin{equation}
\label{eq:Aijkl}
A_{ijkl} = \frac{O_{im}O_{jm}O_{kn}O_{ln}}{G^{(m)}+2G^{(n)}}
\, ,
\end{equation}
where $\TENSOR{O}$ is an orthogonal tensor which diagonalizes $\TENSOR{G}$,
\numparts
\begin{equation}
\label{eq:O}
(\TENSOR{O}\TRANS\TENSOR{G}\TENSOR{O})_{ij} = G^{(i)}\delta_{ij}
\, , \quad
O_{ij}(\TENSOR{O}\TRANS)_{jk} = O_{ij}O_{kj}=\delta_{ik}
\, ,
\end{equation}
such that
\begin{equation}
O_{ij}G_{ik} = O_{ij}G_{ki} = O_{kj}G^{(j)}
\, ,
\end{equation}
\endnumparts
with the eigenvalues $G^{(i)}$ of $\TENSOR{G}$.
Note that the sum in \eref{eq:Aijkl} is over $m$ and $n$ and that
$A_{ijkl}$ obeys the symmetry
$A_{ijkl}=A_{jikl}=A_{ijlk}$.
We then obtain
\begin{equation}
\label{eq:lhs}
J_{\OP{L}_0 A_{ijkl}Y_j Y_k Y_l} = J_{Y_j Y_j Y_i}
- \Theta A_{ijkl} \left( 4G_{jk} J_{Y_l} + 2G_{kl} J_{Y_j} \right)
\, .
\end{equation}
This result corresponds to the left-hand side of \eref{eq:JL0Y}.
For its right-hand side we obtain (see \eref{eq:gL1} and \eref{eq:R1EP})
\begin{eqnarray}
-\int\d\VEC{Y}\,A_{ijkl}Y_j Y_k Y_l \,\OP{L}_1^\dagger\rho_0 =
	(A_{injj}+2A_{ijnj}) \times
\nonumber \\ \qquad\qquad\qquad\qquad \mbox{}
\left(
	\Theta G_{nm} J_{Y_m} -\Theta\frac{\partial\Theta}{\partial X_m}S_{mn}\bar{\rho}_0 + 2\Theta^3 P_n \frac{\partial\bar{\rho}_0}{\partial\F}
\right)
\label{eq:rhs}
\, ,
\end{eqnarray}
where we have used the averages
\numparts
\begin{eqnarray}
\overline{\frac{\partial}{\partial Y_n} Y_j Y_k Y_l} & = &
	\Theta \left( \delta_{nj}\delta_{kl} + \delta_{nk}\delta_{jl} + \delta_{nl}\delta_{jk} \right)
\, , \\
\overline{Y_n Y_j Y_k Y_l} & = &
	\Theta^2 \left( \delta_{nj}\delta_{kl} + \delta_{nk}\delta_{jl} + \delta_{nl}\delta_{jk} \right)
\, , \\
\overline{\VEC{Y}^2 Y_n Y_j Y_k Y_l} & = &
	(n+4)\Theta^3 \left( \delta_{nj}\delta_{kl} + \delta_{nk}\delta_{jl} + \delta_{nl}\delta_{jk} \right)
\, ,
\end{eqnarray}
\endnumparts
and where we have replaced $\Theta V_n\bar{\rho}_0$, emerging from the middle term in the operator $\OP{L}_1^\dagger$, by
$\Theta G_{nm}J_{Y_m} + \Theta \frac{\partial}{\partial X_m} \Theta S_{mn}\bar{\rho}_0 + \left( \Theta^2 Q_n - 2\Theta^3 P_n \right) \frac{\partial\bar{\rho}_0}{\partial\F}$ according to \eref{eq:JYiEPres}.
Combining \eref{eq:lhs} and \eref{eq:rhs} we finally find
\begin{equation}
\label{eq:JYjYjYires}
J_{Y_j Y_j Y_i} =
\Theta (n+2)J_{Y_i}
+ M_{ij} \left( 2\Theta^3 P_j \frac{\partial\bar{\rho}_0}{\partial\F} - \Theta \frac{\partial\Theta}{\partial X_k} S_{kj}\bar{\rho}_0 \right)
\, .
\end{equation}
To arrive at this result we have used
\begin{equation}
(A_{ikjj} + 2A_{ijkj})G_{kl} + (4A_{ijkl} + 2A_{iljk})G_{jk} = (n+2)\delta_{il}
\, ,
\end{equation}
which can be proven using \eref{eq:Aijkl} and \eref{eq:O}
(note that in all terms we sum over $j$ and $k$).
Moreover, we have defined the tensor $\TENSOR{M}$ as
\begin{eqnarray}
M_{ij}
& = & A_{ijkk}+2A_{ikjk} 
\nonumber \\[1ex]
& = & \frac{2 O_{ik}O_{jk}}{3G^{(k)}}+\sum_{l}\frac{O_{ik}O_{jk}}{G^{(k)}+2G^{(l)}}
\nonumber \\
& = & \frac{2}{3}(\TENSOR{G}^{-1})_{ij} + \sum_{l}[(\TENSOR{G}+2G^{(l)}\TENSOR{1})^{-1}]_{ij}
\label{eq:M}
\, .
\end{eqnarray}
We remark that in the case of an isotropic tensor $\TENSOR{G}=g\TENSOR{1}$
(relevant, e.g., for a spherical Brownian particle) $\TENSOR{M}$ simplifies to
$\TENSOR{M} = \frac{(n+2)}{3g}\TENSOR{1}$.

\section{Joint generator for dynamics and functionals}
\label{app:joint}
We here briefly sketch how to obtain the Fokker-Planck equation describing the
evolution of the joint process $\VEC{X}_t,\F_t$ involving the dynamics and the functional.
Note that we do not distinguish here between fast and slow variables, the vector $\VEC{X}$
rather collects all dynamical variables. Likewise, we consider a very general form of the
Langevin equation for $\VEC{X}_t$ and $\F_t$, written component-wise:
\numparts
\begin{eqnarray}
\d X_t^i& = & {u}_i(\VEC{X}_t,,t) \,\d t + \beta_{ij}(\VEC{X}_t,t) \cdot \d W_t^j
\, , \\
d\F_t  & = & {q}_i(\VEC{X}_t,t)\circ dX^i_t + h(\VEC{X}_t,t) \,\d t
\nonumber \\
&\sim& {q}_i(\VEC{X}_t,t)\cdot dX^i_t + \hat{h}(\VEC{X}_t,t) \,\d t
\label{eq:F_ito_st}
\, ,
\end{eqnarray}
\endnumparts
where 
$B_{ij}=\beta_{ik}\beta_{jk}$ and 
$\hat{h}=h+\frac{1}{2}\frac{\partial {q}_i}{\partial X_j}B_{ij}$,
and the symbols $\cdot$ and $\circ$, respectively, 
indicate the It\^o and Stratonovich product.
In going from the first to the second line of \eref{eq:F_ito_st}
we have converted the Stratonovich integral into its statistically equivalent It\^o counterpart.
Similarly to the derivation of It\^o's formula,
we express the differential of a generic function $f(\VEC{X}_t,\F_t)$ as
\begin{eqnarray}
\d f = \left[
 	\d X_i \frac{\partial}{\partial X_i} 
  + \frac{1}{2} \d X_i \d X_j \frac{\partial^2}{\partial X_i\partial X_j}
\right.
\nonumber \\ \qquad\qquad
\left. \mbox{}
	+ \d\F \frac{\partial }{\partial{\F}} + \frac{1}{2}\d\F^2 \frac{\partial^2}{\partial{\F}^2}
	+ \d\F \d X_i \frac{\partial^2}{\partial { \F}\partial X_i}
\right]f
\, .
\end{eqnarray}
Upon taking the average and retaining only the contributions of order ${\cal{O}}(\d t)$
we find
\begin{eqnarray}
 	\frac{\d\langle f \rangle}{\d t}
  = \langle \OP{L} f \rangle
  = \left\langle\left[ u_i\frac{\partial}{\partial X_i} + \frac{1}{2}B_{ij}\frac{\partial^2}{\partial X_i\partial X_j}
\right.\right.\nonumber \\
\qquad\qquad \left.\left. \mbox{}+  
	\left( q_iu_i+\hat{h} \right) \frac{\partial}{\partial{\F}}
  + \frac{1}{2}B_{ij}q_i q_j \frac{\partial^2}{\partial{\F}^2}
  + B_{ij} q_i \frac{\partial^2}{\partial{\F}\partial X_i}
\right] f
\right\rangle
\, ,
\end{eqnarray}
from which we can read off the expression for the joint generator $\OP{L} $.
Its adjoint is
\begin{eqnarray}
\OP{L}^\dagger =
&-& \frac{\partial}{\partial X_i}u_i + \frac{1}{2}\frac{\partial^2}{\partial X_i\partial X_j}B_{ij}
\nonumber\\
&-& \left( q_i u_i+\hat{h} \right) \frac{\partial}{\partial{\F}}
	+ \frac{1}{2}B_{ij} q_i q_j \frac{\partial^2}{\partial{\F}^2}
	+ \frac{\partial^2}{\partial{\F}\partial X_i}B_{ij}q_i
\label{eq:Ldagger}
\, .
\end{eqnarray}
We can now verify that the stochastic differential equations
\eref{eq:dxWNSM}-\eref{eq:siWNSM} given in the main text
correspond to the Fokker-Planck equation \eref{eq:overdampedFPE}.
Let us recall their expression here to ease the procedure:
\numparts
\begin{eqnarray}
\d\VEC{x}_t =
	\frac{1}{\g}(\TENSOR{g}(\VEC{x}_t)\TENSOR{h}(\VEC{x}_t))^{-1} \VEC{f}(\VEC{x}_t,t) \,\d t
  + \VEC{s}(\VEC{x}_t,t) \,\d t
\nonumber \\ \qquad\qquad \mbox{}
	+ \sqrt{2\kB T(\VEC{x}_t,t)/\g}\, [(\TENSOR{g}(\VEC{x}_t)\TENSOR{h}(\VEC{x}_t))^{-1}\TENSOR{\Sigma}(\VEC{x}_t)]^{1/2} \cdot \d\VEC{W}_t
\label{eq:appdxWNSM}
\, , \\[1ex]
\d\F_t = r(\VEC{x}_t,t) \,\d t + \VEC{q}(\VEC{x}_t,t)\circ\d\VEC{x}_t
\label{eq:appdFWNSM}
\, , \\[1ex]
s_i(\VEC{x},t) =
	\frac{\kB T(\VEC{x},t)}{\g}
	\frac{\partial[(\TENSOR{g}(\VEC{x})\TENSOR{h}(\VEC{x}))^{-1}]_{ij}}{\partial x_k} \Sigma_{jk}
\label{eq:appsiWNSM}
\, ,
\end{eqnarray}
\endnumparts
By plugging the expressions from \eref{eq:appdxWNSM}-\eref{eq:appsiWNSM}, which
correspond to $u_i$, $q_i$, $\hat{h}$ and $B_{ij}$, into \eref{eq:Ldagger},
dropping the explicit space and time dependence we obtain
\begin{eqnarray} 
\frac{\partial\rho}{\partial t}
= \OP{L}^\dagger \rho
= &-&\frac{\partial}{\partial x_i}
	\left(
	\frac{1}{\g}[(\TENSOR{g}\TENSOR{h})^{-1}]_{ij} f_j
  + \frac{\kB T}{\g} {\Sigma}_{jk} 
  	\frac{\partial [(\TENSOR{g}\TENSOR{h})^{-1}]_{ij}}{\partial x_k}
	\right)\rho
\nonumber \\ \mbox{}
  &+& \frac{1}{2} \frac{\partial^2}{\partial x_i \partial x_j}
	\left(
	\frac{2\kB T}{\g}\, [(\TENSOR{g}\TENSOR{h})^{-1}\TENSOR{\Sigma}]_{ij}
	\right)\rho
\nonumber \\ \mbox{}
  &-& \Bigg[
  	q_i \left( \frac{1}{\g}[(\TENSOR{g}\TENSOR{h})^{-1}]_{ij} f_j     
  + \frac{\kB T}{\g} {\Sigma}_{jk} \frac{\partial [(\TENSOR{g}\TENSOR{h})^{-1}]_{ij}}{\partial x_k}\right)
\nonumber \\
  && \qquad\qquad \mbox{}+ \frac{1}{2}\frac{\partial {q}_i}{\partial x_j}
  	\left( \frac{2\kB T}{\g}\, [(\TENSOR{g}\TENSOR{h})^{-1}\TENSOR{\Sigma}]_{ij}\right)+r 
\Bigg] \frac{\partial \rho}{\partial{\F}}
\nonumber \\ \mbox{}
  &+& \frac{1}{2}
  \left(
  	\frac{2\kB T}{\g} \, [(\TENSOR{g}\TENSOR{h})^{-1}\TENSOR{\Sigma}]_{ij}
  \right) q_i q_j \frac{\partial^2 \rho}{\partial{\F}^2}
\nonumber \\ \mbox{}
  &+& \frac{\partial}{\partial x_i}
  \left(
  	\frac{2\kB T}{\g} \, [(\TENSOR{g}\TENSOR{h})^{-1}\TENSOR{\Sigma}]_{ij}
  \right)q_j \frac{\partial \rho}{\partial{\F}}
\, .
\end{eqnarray}
In dimensionless form,  if $(\TENSOR{g}\TENSOR{h})^{-1}\Sigma$ is symmetric, which is guaranteed if the fast process obeys detailed balance,  this is equivalent to 
\begin{eqnarray}
\frac{\partial\rho}{\partial\tilde{t}} =
-\left[
	\frac{\partial}{\partial\tilde{x}_i} [(\TENSOR{g}\TENSOR{h})^{-1}]_{ij} \tilde{f}_j
	- \frac{\partial}{\partial\tilde{x}_i} [(\TENSOR{g}\TENSOR{h})^{-1}]_{ij} \frac{\partial}{\partial\tilde{x}_k} \Sigma_{jk}\Theta
\right.
\nonumber \\ \qquad\qquad \mbox{}
	+ \tilde{r} \frac{\partial}{\partial\tilde{\F}}
	+ \tilde{q}_i [(\TENSOR{g}\TENSOR{h})^{-1}]_{ij} \tilde{f}_j \frac{\partial}{\partial\tilde{\F}}
	+ \Theta \frac{\partial \tilde{q}_i [(\TENSOR{g}\TENSOR{h})^{-1}]_{ij}}{\partial\tilde{x}_k}
		\Sigma_{jk} \frac{\partial}{\partial\tilde{\F}}
\nonumber \\ \qquad\qquad \mbox{}
	- \frac{\partial}{\partial\tilde{x}_i} \Theta [(\TENSOR{g}\TENSOR{h})^{-1}]_{ij} \Sigma_{jk} \tilde{q}_k
		\frac{\partial}{\partial\tilde{\F}}
	- \frac{\partial}{\partial\tilde{x}_k} \Theta \tilde{q}_i [(\TENSOR{g}\TENSOR{h})^{-1}]_{ij} \Sigma_{jk}
		\frac{\partial}{\partial\tilde{\F}}
\nonumber \\ \qquad\qquad\qquad\qquad\qquad\qquad\qquad\quad \mbox{}
\left.
	- \tilde{q}_i [(\TENSOR{g}\TENSOR{h})^{-1}]_{ij} \Sigma_{jk} \tilde{q}_k \Theta \frac{\partial^2}{\partial\tilde{\F}^2}
\right] \rho
\, ,
\end{eqnarray}
which is exactly the Fokker-Planck equation given in \eref{eq:overdampedFPE}.

{\color{white}x}\\[5ex]
\bibliographystyle{iopart-num}

\providecommand{\newblock}{}

\end{document}